\documentclass{aa520}  

\usepackage{graphicx}
\usepackage{natbib}

\usepackage{txfonts}
\usepackage{amsfonts}
\usepackage{booktabs}
\usepackage{siunitx}
\usepackage{multirow}
\usepackage[version=4]{mhchem}
\usepackage{newtxtext,newtxmath}
\usepackage{tikz,xcolor,hyperref}
\definecolor{lime}{HTML}{A6CE39}
\DeclareRobustCommand{\orcidicon}{%
	\begin{tikzpicture}
	\draw[lime, fill=lime] (0,0) 
	circle [radius=0.16] 
	node[white] {{\fontfamily{qag}\selectfont \tiny ID}};
	\draw[white, fill=white] (-0.0625,0.095) 
	circle [radius=0.007];
	\end{tikzpicture}
	\hspace{-2mm}
}

\foreach \x in {A, ..., Z}{%
	\expandafter\xdef\csname orcid\x\endcsname{\noexpand\href{https://orcid.org/\csname orcidauthor\x\endcsname}{\noexpand\orcidicon}}
}

\begin{document}

   \title{Virgo Filaments. III. The gas content of galaxies in filaments as predicted by the GAEA semi-analytic model}

\titlerunning{Gas content of galaxies in filaments as predicted by GAEA}

\authorrunning{D. Zakharova et al.}

   \author{D. Zakharova\inst{1,2}\orcidD{},
          B. Vulcani\inst{2}\orcidB{}, 
          G. De Lucia\inst{3,4}\orcidC{},
          R. A. Finn\inst{5},\orcidR{}
          G. Rudnick\inst{6}\orcidG{},
          F. Combes\inst{7}\orcidX{},
          G. Castignani\inst{8}\orcidQ{},
          F. Fontanot\inst{3,4}\orcidF{},
          P. Jablonka\inst{9}\orcidP{},
          L. Xie\inst{10,3}\orcidL{},
          M. Hirschmann\inst{11,3}\orcidM{}
             }
    \institute{Dipartimento di Fisica e Astronomia Galileo Galilei, Universit\`a degli studi di Padova, Vicolo dell’Osservatorio, 3, I-35122 Padova, Italy
    \and
    INAF – Osservatorio astronomico di Padova, Vicolo dell’Osservatorio, 5, I-35122 Padova, Italy
    \and
    INAF – Osservatorio Astronomico di Trieste, Via Tiepolo 11, I-34131 Trieste, Italy
    \and
    IFPU - Institute for Fundamental Physics of the Universe, via Beirut 2, 34151, Trieste, Italy
    \and
    Siena College, 515 Loudon Rd., Loudonville, NY 12211
    \and
    The University of Kansas, Department of Physics and Astronomy, Malott Room 1082, 1251 Wescoe Hall Drive, Lawrence, KS, 66045, USA
    \and
    Observatoire de Paris, LERMA, Coll`ege de France, CNRS, PSL University, Sorbonne University, 75014, Paris
    \and
    INAF - Osservatorio di Astrofisica e Scienza dello Spazio di Bologna, via Gobetti 93/3, I-40129, Bologna, Italy
    \and
    Laboratoire d’astrophysique, Ecole Polytechnique Fédérale de Lausanne (EPFL), Observatoire, CH-1290 Versoix, Switzerland
    \and
    Tianjin Normal University, Binshuixidao 393, 300387, Tianjin, China
    \and
    Institute for Physics, Laboratory for Galaxy Evolution and Spectral modelling, Ecole Polytechnique Federale de Lausanne,
    Observatoire de Sauverny, Chemin Pegasi 51, 1290 Versoix, Switzerland
    }
    
   \date{Received May 22, 2024; accepted August 06, 2024}

 
  \abstract
   {Galaxy evolution depends on the environment in which galaxies are located. The various physical processes (ram-pressure stripping, tidal interactions, etc.) that are able to affect the gas content in galaxies have different efficiencies in different environments. In this work, we examine the gas (atomic \ce{HI} and molecular \ce{H2}) content of local galaxies inside and outside clusters, groups, and filaments as well as in isolation using a combination of observational and simulated data. We exploited a catalogue of galaxies in the Virgo cluster~(including the surrounding filaments and groups) and compared the data against the predictions of the Galaxy Evolution and Assembly (GAEA) semi-analytic model, which has explicit prescriptions for partitioning the cold gas content in its atomic and molecular phases. We extracted from the model both a mock catalogue that mimics the observational biases and one not tailored to observations in order to study the impact of
observational limits on the results and predict trends in regimes
not covered by the current observations. The observations and simulated data show that galaxies within filaments exhibit intermediate cold gas content between galaxies in clusters and in isolation. The amount of \ce{HI} is typically more sensitive to the environment than \ce{H2} and low-mass galaxies ($\log_{10} [{\rm M}_{\star} / \rm{M}_{\sun} ] < 10$) are typically more affected than their massive ($\log_{10} [{\rm M}_{\star} / \rm{M}_{\sun} ] > 10$) counterparts. Considering only model data, we identified two distinct populations among filament galaxies present in similar proportions: those simultaneously lying in groups and isolated galaxies. The former has properties more similar to cluster and group galaxies, and the latter is more similar to those of field galaxies. We therefore did not detect the strong effects of filaments themselves on the gas content of galaxies, and we ascribe the results to the presence of groups in filaments.}


   \maketitle
%
\section{Introduction}
\label{section:intro}

In recent years, studies on the influence of the environment have established that the properties of galaxies correlate with their local density. Galaxies in clusters  typically have earlier morphological types~\citep{Dressler+1980_morph, Vulcani+2023a}, are more massive, ~\citep{Kauffmann+2004, Baldry+2006}, have reduced star formation~\citep{Peng+2010, Woo+2013, Vulcani+2010, Paccagnella+2016, Finn+2023}, and contain less cold gas~\citep{Giovanelli+1985, Brown+2017} than galaxies in lower density regions~\citep{Rojas+2004, Beygu+2016}.
\par
While the most striking differences are found when comparing galaxies in the field and in clusters, it has become clear that `intermediate' environments, such as groups and filaments, also play an important role. Nevertheless, determining the environment of a galaxy poses a challenge. For instance, there is no obvious separation between clusters and groups, as they can be considered as elements of the large-scale structure~(LSS) gathering into filaments and walls~\citep{Bond+1996}. Secondly, filament identification is challenged by a number of observational effects (e.g. the fingers-of-god effect due to the peculiar velocities of galaxies) that distort the actual distribution of galaxies~\citep{Kuchner+2021}, which in turn affects filament extraction. Equally important, the identification of filaments depends on the tracer: \cite{Zakharova+2023} have shown that galaxies of different mass trace the underlying distribution of dark matter differently.
\par
Despite the difficulties related to the determination of the galaxy environments, it has been found that galaxies within filaments exhibit notable differences from their cluster and field counterparts. Independent studies consistently indicate that filament galaxies tend to be more massive~\citep{Laigle+2017,Malavasi+2017,Zakharova+2023} and redder~\citep{Kuutma+2017,Kraljic+2018, Singh+2020} and to have lower star-formation rates \citep[SFRs;][]{Kraljic+2018,Sarron+2019} than galaxies in the field.
Some studies have found evidence of a distinct impact of filaments on different gas phases. \cite{Vulcani+2019} found that in some filament galaxies, ionised H$\alpha$ clouds extend far beyond what is seen for other non-cluster galaxies. This result may be due to the effective cooling of the dense star-forming regions in filament galaxies, which can increase the spatial extent of the H$\alpha$ emission. Also, atomic \ce{HI} and molecular gas reservoirs have been shown to be impacted by the filament environment~\citep{Kleiner+2017, Odekon+2018, Blue_Bird+2020, Lee+2021, Castignani+2022_gas}.

The effect of filament environment on galaxy properties remains debated because filaments contain groups of various masses~\citep{Tempel+2014_perls} that may contribute to the measured differences of the properties of filament members with respect to those of galaxies in the field ~\citep{Sarron+2019}.
\par
The trends observed for galaxies in dense regions can be explained by the impact of mechanisms characteristic of dense environments, such as ram-pressure stripping of gas~\citep{GunnGott1972, Bahe+2017}, tidal effects~\citep{Bekki+1998}, galaxy-galaxy interactions~\citep{Naab+2007}, and mergers~\citep{Mihos1996, Kaviraj+2009}. These processes typically affect the gas content of galaxies since they can displace and remove it, resulting in the suppression of star formation~\citep{De_Lucia+2012, Wetzel+2012}. All of the mechanisms listed above affect the gas content of galaxies. Therefore, a strong correlation is expected between the amount of gas in galaxies and their environment. 
\par
The nearby massive Virgo cluster and its surrounding filaments are an ideal laboratory for comparing the properties of galaxies in various environments. 
The first work of this series,~\citep{Castignani+2022_gas}, focused on gathering and analysing data about the gas content of galaxies around the Virgo cluster. The authors of that work collected both atomic and molecular gas content information for galaxies within the cluster~(from \citealt{Boselli+2014}) and filaments in an extended region around the cluster by using both new observations and existing data in the literature. Data were compared with isolated galaxies from the AMIGA survey~\citep{AMIGA}, and the results showed a decreasing
gas content moving from the field to filaments and then clusters
as well as an increase of the proportion of quiescent galaxies. \cite{Castignani+2022_gas} concluded that the processes leading to the suppression of star formation in galaxy clusters are already starting to take place in filaments.
\par
The second paper of this series, \citep{Castignani+2022_catalogue}, compiled an exceptional dataset for $\sim$7000 galaxies around the Virgo cluster into a catalogue~(based on the \citealt{Kim+2014} catalogue), combining spectroscopically confirmed sources across multiple databases and surveys, such as HyperLeda, NASA Sloan Atlas, NED, and ALFALFA. The resulting catalogue provides positions, masses, integrated \ce{HI} and \ce{CO}, and a parametrisation of the environment for galaxies surrounding Virgo.  In addition, \cite{Castignani+2022_catalogue} conducted an analysis of galaxy properties within Virgo filaments, confirming that filament members indeed have intermediate properties~(local density, galaxy morphology, bar fractions) between galaxies in the cluster and the field. 
\par
\textcolor{black}{
This paper is the third of the series, and it is dedicated to examining two main points. First, we wanted to test whether current state-of-the-art semi-analytical models can reproduce the observed gas properties of galaxies across different environments. Second, we wished to investigate how the gas content of galaxies around the Virgo cluster depends on the environment. In particular, our aim is to understand the role of filaments in regulating the gas content of galaxies from a theoretical point of view. To do so, we took advantage of a state-of-the-art theoretical model of galaxy formation, GAEA.  Unlike widely used hydrodynamical models such as IllustrisTNG~\citep{Illustris1, Illustris2,Illustris3,Illustris4, Illustris5}, EAGLE~\citep{EAGLE_1,EAGLE_2}, and The Three Hundred project~\citep{TheThreeHundredproject} or constrained simulations such as Simulating the LOcal Web~\citep[SLOW;][]{SLOW1, SLOW2,SLOW3} and HESTIA~\citep{HESTIA}, GAEA includes an explicit treatment for the partition of cold gas in its atomic and molecular components, and it is coupled to a large cosmological volume with relatively high resolution.   } 
\par
The paper is organised as follows. \hyperref[section:data]{Section 2}  describes the observational and model samples used in this work. 
In \hyperref[sec:environments]{Sect. 3}, we parametrize the environment for each observed and model galaxy. In \hyperref[sec:finalisingdata]{Sect. 4}, we describe how we constructed our model mock sample to be compared with data. \hyperref[sec:results]{Section 5} compares the gas properties of galaxies in the cluster, filaments, and field both for the observational sample and the mock data.  In \hyperref[sec:discussion]{Sect. 6}, we discuss the role of filaments in regulating the gas content in the Virgo cluster surroundings. \hyperref[sec:conclusion]{Section 7} summarises the results of this paper.

\section{Data description}
\label{section:data}

\subsection{Observational data}
\label{subsec:obs_data}

We made use of the Virgo Filament catalogue, which was released in \cite{Castignani+2022_catalogue}. The catalogue is based on data from different databases and surveys, including HyperLeda, NASA Sloan Atlas, and NASA/IPAC Extragalactic Database, and contains information about 6780 galaxies within $\sim$12 virial radii around the Virgo cluster. The catalogue covers the region delimited by $100^{\circ} < RA < 280^{\circ}$ and  $-1.3^{\circ} < DEC < 75^{\circ}$,  and contains galaxies with recessional velocities $500 < v_r < 3300$ km/s. It also includes 110 galaxies that have recessional velocities $<500$ km/s but have redshift-independent distances in the NASA/IPAC Extragalactic Database  (NED-D; \citealt{Steer+2017}) that correspond to cosmological velocities in the range of 500–3300 km/s. Some of these galaxies are Virgo cluster members that are located near the caustics, and thus have the largest deviations in velocity from Virgo. More details on the catalogue construction and on how it was cleaned from spurious sources, stars, and duplicates can be found in \cite{Castignani+2022_catalogue}.
\par

We estimated the stellar masses and SFRs from spectral energy distribution (SED) fitting.  We construct the SEDs from publicly available, wide-area imaging surveys that span from the UV to the infrared.  Specifically, we use: FUV and NUV from GALEX~\citep{Gil_de_Paz+2007}; $grz$ imaging from the DESI Legacy Imaging Surveys \citep{DESI+2016}; and 3.4$\mu$, 4.5$\mu$, 12$\mu$  and 22$\mu$ \ from WISE~\cite{Wright2010}.  Magnitudes in each photometric band are determined from a custom elliptical aperture photometry pipeline that is optimised for large, nearby galaxies. The photometry and masking methods are based on those developed for the {Siena Galaxy Atlas} and are described in detail in \citet{Moustakas+2023}. 
Our fluxes are measured within a fixed  elliptical
aperture whose semi-major axis is 1.5 times the
estimated size of the galaxy based on the second moment of the light
distribution (after subtracting stars and masking out surrounding
galaxies in the image).  We do not attempt to correct the aperture fluxes to total fluxes.  However, using a curve-of-growth analysis, we estimate that the correction would affect the stellar masses by $<20$\%. To correct for galactic extinction, we use the redenning values from \citet{Schlegel1998} and follow the Legacy Survey's procedure to transform to the $grz$ and WISE filters.  We transform $E(B-V)$ to extinction in the GALEX FUV and NUV filters using the transformations in \citet{Wyder2007}. 

We used the Multi-wavelength Analysis of Galaxy Physical Properties (MAGPHYS) tool \citep{daCunha+2008} to model the SEDs and estimate  stellar masses and SFRs~(rely on the \cite{Chabrier+2003_imf} initial mass function).  Following the Legacy Survey, we use the $gz$ filters for galaxies with Declination $\delta > 32.375$ and the $grz$ filters for galaxies south of this limit.  This difference in the inclusion of the $z$-band is because there are known offsets in the relative $z$-band calibration in the northern survey that are more pronounced for galaxies with larger angular extent.  We verified using the southern filters that removing the $z$-band did not affect our SED-fitting results. The southern filters were already incorporated into MAGPHYS, and the northern  filters were added to the MAGPHYS package following a request to its creator (Da Cunha, private communication). 

We determined the stellar mass completeness limit, above which we will detect all galaxies regardless of their $r$-band stellar mass-to-light ratio (${\rm M}_\star/L_r$).  We derived the stellar mass completeness limit using a technique adapted from \citet{Marchesini+2009}, \citet{Rudnick+2017}, and \citet{Finn+2023}.  We started with galaxies at the high velocity (distance) end of our survey, namely those with $2500 < v/{\rm km/s} < 3500$ as these will have the faintest observed brightness at a fixed mass or luminosity.  We selected all galaxies between 0.5 and 1.25 mag brighter than the SDSS spectroscopic limit of $m_r = 17.77$.  These galaxies are bright enough that we should be able to detect all galaxies with equal completeness, regardless of their ${\rm M}_\star/L_r$.   We make the reasonable assumption that ${\rm M}_\star/L_r$ does not vary strongly with observed magnitude over this range.  Therefore, the distribution of ${\rm M}_\star/L_r$ for this bright subsample should be consistent with the intrinsic ${\rm M}_\star/L_r$ distribution near our apparent magnitude limit.  Using this distribution of ${\rm M}_\star/L_r$, we measure the 5\% highest ${\rm M}_\star/L_r$.  At the luminosity limit of our survey, corresponding to the apparent magnitude limit of the most distant galaxies, this ${\rm M}_\star/L_r$ limit yields a stellar mass limit of $\log({\rm M}_\star / {\rm M}_\odot) = 8.26.$  Galaxies at lower stellar masses would only be detectable if they had lower ${\rm M}_\star/L_r$ values.

From here on, we use only galaxies with the measured stellar masses 
above the mass completeness limit of the catalogue~($ \log [ {\rm M}_{\star} / \rm{M}_{\sun}] > 8.3$), for a total of  2919 galaxies. We used this sample to identify filaments (see  Sect.~\ref{sec:environments}).

For part of the \cite{Castignani+2022_catalogue} sample, \ce{HI} and \ce{H2} observations were obtained in \cite{Castignani+2022_gas}. They presented a compilation of the available data: 
the catalogue contains information about  atomic ($\rm{M}_{\ce{HI}}$) and molecular hydrogen ($\rm{M}_{\ce{H2}}$) for galaxies with stellar masses  $\rm 9 < \log_{10}({M}_{\star}/M_\sun) < 11$.  
Specifically, data are available for 389 galaxies of which 97 are cluster galaxies, 166 filament galaxies\footnote{We note that \cite{Castignani+2022_catalogue} used a different approach to identify filament members, but this does not affect the results (sec \ref{subsec:filaments_idn}).}, and 132 are galaxies in the field. Briefly, \cite{Castignani+2022_gas} collected \ce{HI} observations for the Virgo cluster galaxies from \cite{Boselli+2014}. For the galaxies in the longest filaments with the highest contrast around Virgo,  they collected data from the literature and observed the missing galaxies with the Nançay telescope~(59 galaxies in the catalogue).  \ce{HI} masses for field galaxies were also taken from the literature (mainly \citealt{AMIGA, Springob+2005}). 82 galaxies with \ce{HI} measurements had molecular hydrogen measurements from the literature, while the rest were observed with the IRAM-30m~(both \ce{CO}(1 $\to$ 0) and \ce{CO}(2 $\to$ 1), simultaneously). A detailed description of these data and how \ce{HI} and \ce{H2} masses were estimated can be found in \cite{Castignani+2022_gas}. 

\subsection{Simulated data}
\label{subsec:gaea_data}

We used predictions from the GAlaxy Evolution and Assembly (GAEA) semi-analytic model~\citep{Hirschman+2016} coupled with the Millennium II Simulation \citep{Boylan-Kolchin_etal}. GAEA\footnote{Information about the GAEA model and selected model predictions can be found at  https://sites.google.com/inaf.it/gaea} builds on the original model presented in \citet{DeLucia_and_Blaizot}, but it includes a number of important updates. In particular, we use here the latest rendition of the model presented in \citet{De+Lucia+2024}, that includes: (i) a detailed treatment for the non-instantaneous recycling of gas, metals, and energy that allows different elemental abundances to be traced explicitly \citep{DeLucia+2014}; (ii) an updated treatment for stellar-feedback that provides  good agreement with the observed evolution of the galaxy stellar mass function up to $z\sim 3$ and other important scaling relations \citep{Hirschman+2016}; (iii) an explicit treatment for partitioning the cold gas in its atomic and molecular components, and for ram-pressure stripping of both the hot gas and cold gas reservoirs of satellite galaxies \citep{Xie+2017,Xie+2020}; (iv) an updated treatment of AGN feedback that includes an improved modelling of cold gas accretion on supermassive black holes and an explicit implementation for quasar winds \citep{Fontanot_etal_2020}.  \citet{De+Lucia+2024} have shown that the latest renditions of GAEA provides an improved agreement with the observed distributions of specific SFRs in the local Universe, as well as a quite good agreement with the observed passive fractions up to $z\sim 3$, making this model version an ideal tool to interpret the data considered in this work. The results of the model are based on the Chabrier IMF~\cite{Chabrier+2003_imf}.

We took advantage of a GAEA realisation run on dark matter halo merger trees extracted from the Millennium II simulation, which followed 
2,160$^3$ dark matter
particles in a box of 100~Mpc$\,{h}^{-1}$ on a side , with cosmological parameters consistent with WMAP1 ($\Omega_\Lambda=0.75$,
$\Omega_m=0.25$, $\Omega_b=0.045$, $n=1$, $\sigma_8=0.9$, and $H_0=73 \, {\rm km\,s^{-1}\,Mpc^{-1}}$). The high resolution of the simulation (the particle mass is $6.9\times10^6\,{\rm M_{\sun}}\,{\rm h}^{-1}$) allows galaxies to be well resolved down to stellar masses of $10^8\,{\rm M}_{\sun}$.
\par
For our analysis, we used the following information for each model galaxy: 3D positions and velocities, the mass of the hosting dark matter halo~($\rm{M}_{200}$,  mass of the region enclosing a mean density of 200$\rho_{\rm crit}$, where  $\rho_{\rm crit}$ is critical density of the Universe), stellar mass,  mass of \ce{HI} and of \ce{H2}, galaxy type  (central or satellite), and star formation rate. 
\par
The model includes an explicit treatment of the interaction of satellite galaxies with the host halo gas~(both stripping of the hot gas associated with satellites and ram-pressure stripping of cold gas). A detailed description of that can be found in \cite{Xie+2020}. Being coupled with merger trees extracted from N-body simulations, it also accounts for assembly bias~-- that is, earlier assembled haloes are more clustered than later assemblies of similar mass~\citep{Gao+2005}, which leads to an impact on galaxies properties~\citep{Croton+2007,Wang+2013}. We emphasize that the GAEA model does not include any explicit mechanism accounting for the interaction of galaxies with filaments.
\par
For comparison with the observations, we extract from GAEA all halos having a mass similar to that of Virgo~($M_{\rm Virgo} = 4.5 \cdot 10^{14} \rm{M}_{\sun}$ \citealt{Kourkchi+2017}) \textcolor{black}{at $z\sim0$}. Only three such halos exist in the Millennium II volume, and their virial masses are $4.7 \cdot 10^{14}\,{\rm M}_{\sun}$~(GAEA V1), $4 \cdot 10^{14}\,{\rm M}_{\sun}$~(GAEA V2) and $4.9 \cdot 10^{14}\,{\rm M}_{\sun}$~(GAEA V3). 

\begin{figure*}
    \centering
    \includegraphics[width=1\linewidth]{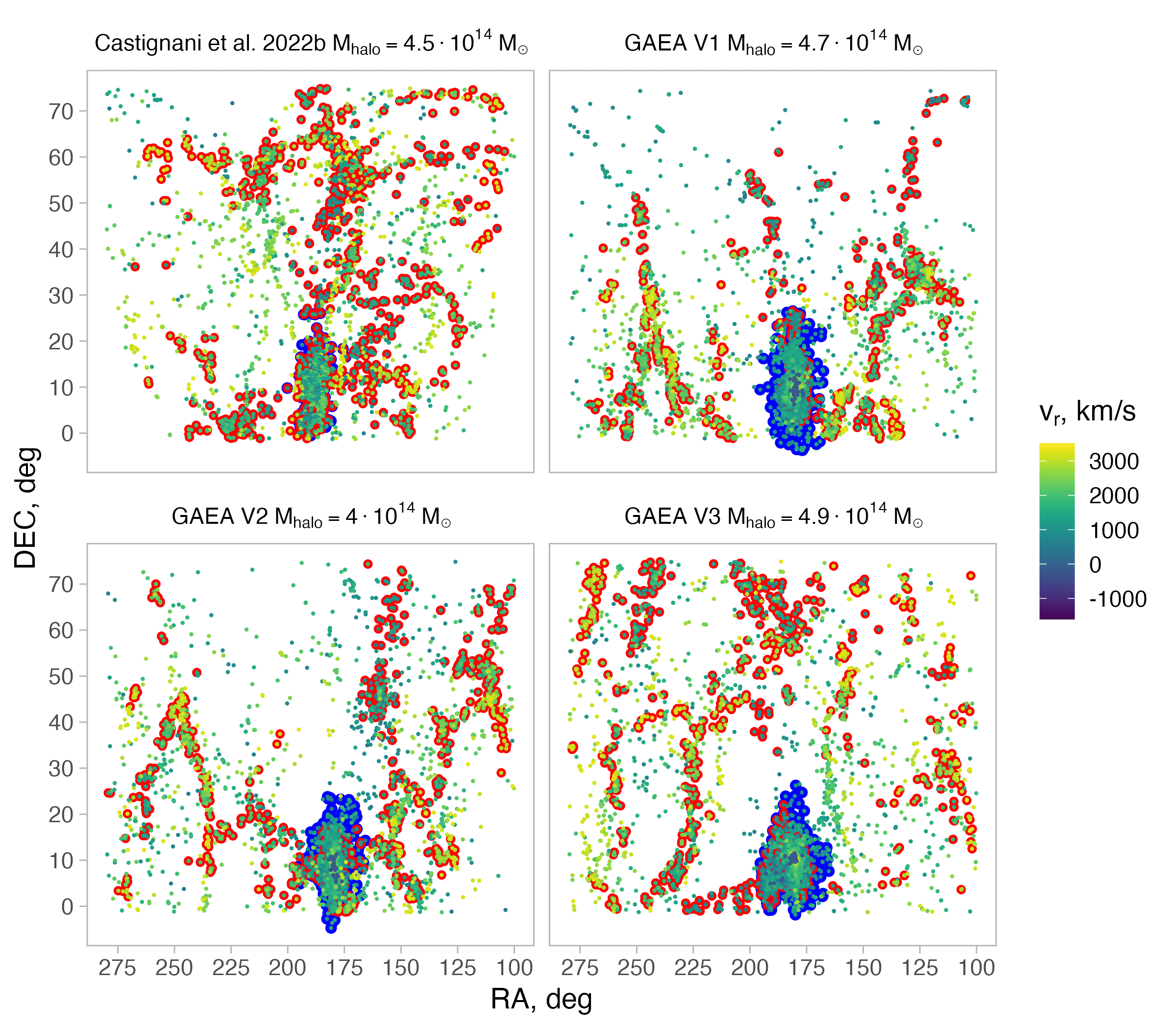}
    \caption{Distribution of galaxies around the Virgo cluster~(top-left panel, from \citealt{Castignani+2022_catalogue}) and  around the three Virgo-like clusters in the GAEA model~(top right: GAEA V1, bottom left: GAEA V2, bottom right: GAEA V3) in celestial coordinates~(corresponding to the GAEA-all sets; Sect.~\ref{sec:gasmass_incomplet_cuts}).  The label of each panel indicates the mass of  Virgo or of the Virgo-like halos. Each point is a galaxy with  $\rm{M}_{\star} > 10^{8.3} \rm{M}_{\sun}$~(mass completeness limit) and is colour-coded by its recessional velocity. Additionally, galaxies belonging to filaments (see Sect.~\ref{subsec:filaments_idn}) are highlighted in red\textcolor{black}{, and the cluster members in blue.} }
    \label{fig:radec_filsmems}
\end{figure*}

\subsubsection{GAEA coordinate transformation}
\label{subsec:gaea_coord_transform}
The first step of our analysis is to extract from the model simulated box portions of the sky of a size comparable to the one covered by the observations.  The catalogue by \citet{Castignani+2022_catalogue} is  based on a RA-DEC-$v_{r}$ selection within a fixed area around Virgo. We obtain the same coordinates for each galaxy in GAEA: first of all, we position the simulated volume at the same distance of the Virgo cluster ($\sim16$ Mpc/h, \citealt{Mei+2007}) and select galaxies in a range of radial velocities $v_r$ relative to the cluster centre similar to that of the observational sample. Next, we transform the  GAEA cartesian coordinates x-y-z in RA-DEC-$v_{r}$ to be able to cut the same region in RA-DEC-$v_{r}$ coordinates. We also obtain supergalactic coordinates SGX-SGY-SGZ, which we use to identify filaments. This procedure is detailed in Appendix~\ref{app:coord_transform}. 

As a final step, we select a region similar to the one analysed by \cite{Castignani+2022_catalogue} around the Virgo cluster and  consider only galaxies with 
$100^{\circ} < RA < 280^{\circ}$ and  $-1.3^{\circ} < DEC < 75^{\circ}$ and have matching velocities $500 < v_r < 3300$ km/s.  Since some of the Virgo cluster members have $v_r < 500$ km/s (see \citealt{Castignani+2022_catalogue} for more details), then we do not apply this condition for cluster members~(galaxies inside 3 virial radii of the cluster centre, regardless of their $v_r$). We hence obtain 
three regions of the sky around three Virgo-like systems~(GAEA V1, GAEA V2, and GAEA V3). 
We note that our cubes include distortions in the distribution of galaxies associated with line-of-sight effects~(as do the observational data). We do not make any adjustments to these effects to be consistent with the previous works in the series.

\subsubsection{Stellar mass completeness limit and the selection function}
\label{subsec:masscut_83}
Next, we apply the same mass cut estimated for the observations ($ \log {\rm M}_{\star}/{\rm M}_{\sun} > 8.3$). At this stage, the number of galaxies in each of the three GAEA regions is on average  3-5 times higher than the sample of \cite{Castignani+2022_catalogue} above the same mass completeness cut. As discussed above, the Virgo Filament catalogue is a combination of different datasets and, as such, it is characterised by a selection function that is hard to precisely replicate \citep{Castignani+2022_catalogue}. 
Instead of trying to emulate the `incompleteness' for each of the three regions, we reduce the number of galaxies by performing one random extraction of a sample that has the same number of galaxies found in the observed sample (2518) and a similar stellar mass distribution.  
\par

Figure~\ref{fig:radec_filsmems} shows the result of the extraction of three Virgo-like regions and the observations in the RA-DEC plane.

\section{Environmental definitions}
\label{sec:environments}
\subsection{Identification of filaments}
\label{subsec:filaments_idn}

To identify filaments, \cite{Castignani+2022_catalogue} exploited a tomographic approach to characterize  the highest density contrasts relative to the surrounding field as determined by visual inspection. Briefly, they considered the eight filamentary structures presented in \citet{Tully+1982} and \citet{Kim+2016}, and visually identified 5 additional filaments. For each filament, they considered an associated cuboid in the 3D supergalactic frame large enough to enclose all galaxies that belong to the structure under consideration. They then determined the filament spines by fitting the locations of the galaxies in supergalactic coordinates. The method developed by \cite{Castignani+2022_catalogue} requires visual inspection, which makes it very difficult to replicate. Therefore, we opt for a redefinition of the filamentary structure based on the Discrete Persistent Structures Extractor (DisPerSE, \citealt{Sousbie+2011, Sousbie_etal+2011}) code. In this way, we can rely on a consistent definition between the observational and simulated samples. We refer to the original papers for detailed information on the algorithm employed by DisPerSE. Briefly, using information about the distribution of galaxies, the code estimates a density distribution that is then used to identify the spines of the filamentary structure. Different `persistence' levels can be chosen to identify filaments with different contrast. The higher the persistence level, the higher the density of the detected filaments: for instance, a threshold of $7\sigma$ finds only the densest structures, while a threshold of $3\sigma$ finds many more short~filaments (length less than a typical cluster radius), many of which might correspond to spurious detections~(see \citealt{Zakharova+2023} for more details). 
\par
We tested   using  DisPerSE on the observational sample and we recover approximately the same structures identified  by \cite{Castignani+2022_catalogue}, although with different levels of details. Using the supergalactic coordinates SGX-SGY-SGZ, we extracted filaments from the observed sample adopting different persistence levels ($3\sigma$, $4\sigma$, $5\sigma$, and $7\sigma$)  to identify filaments characterised by different density contrasts.
\par
At the $3\sigma$ threshold, the DisPerSE-defined filaments system~(FS) catches almost all the structures  defined by  \cite{Castignani+2022_catalogue} but also a number of additional filaments. Adopting this persistence level, up to 70\%  of the galaxies turn out to be in filaments. This values is too large when compared to the catalogue by  \cite{Castignani+2022_catalogue}. In addition, many of the identified filaments  are  extremely faint. In contrast, the FS obtained using a persistence level of $5\sigma$ or larger loses some of the filaments identified by \cite{Castignani+2022_catalogue}, including some very dense ones. 
As a compromise, we choose a $4\sigma$ persistence level.  With this threshold, we find that the visual approach and DisPerSe would give consistent 
result.
We have verified that the adoption of a new method to determine  the filaments has no impact on the scientific results obtained with the  approach by \cite{Castignani+2022_catalogue}. 
\par
We apply the $4\sigma$  persistence level for the extraction of filaments using DisPerSE both for observations and for the simulated regions. In both cases, we also remove all filaments that are shorter than 3 Mpc/h~(of the order of 7$\pm$3 filaments depending on the analysed sample), as  it is hard to establish if they are real structures. As noted above, the  simulated regions GAEA V1, GAEA V2, and GAEA V3 described in Sect.~\ref{subsec:gaea_coord_transform} include the fingers-of-god effect (the elongation along the line-of-sight). This also affects filament identification, as we discuss  in detail in Appendix~\ref{appendix:impact_of_elongation}. Briefly, elongation along line-of-sight does not greatly interfere with the classification of galaxies as members in filaments, but it also does not allow one to determine the exact distance to the axis of the filaments. 
\par
Following \cite{Castignani+2022_catalogue}, in both the model and observations,  we consider a galaxy to be in a filament if its distance to the nearest filament segment is less than 2 Mpc/h~ and if the galaxy does not belong to the cluster (see the cluster membership definition below). We exclude from the filaments the cluster members as we expect that the effect of the cluster environment is dominant over the possible effect of the filaments~\citep{Sarron+2019,Kraljic+2018}. 
Figure \ref{fig:radec_filsmems} also highlights in red the members of the selected filaments for the three extracted clusters and the observed data.

\subsection{Additional environments around Virgo}
\label{subsec:env_idn}

In addition to the filaments, \cite{Castignani+2022_catalogue} considered other environments in the region around the Virgo cluster. First of all, they identified cluster members, selecting galaxies within 3.6 Mpc/h from the Virgo cluster centre in the 3D SG coordinate frame. The chosen radius corresponds approximately to $\sim$3$R_{200}$. They also considered as cluster members those galaxies that fall within the cluster region delimited by the caustics in the phase-space diagram, regardless of their  SG coordinates.
Then, they  identified galaxy groups by matching their  catalogue to the environmental catalogue from \cite{Kourkchi+2017}, who characterised galaxy groups in our immediate neighbourhood ($v_r$ $<$ 3500 km/s). 
Finally, they assembled a sample of pure field galaxies, that is, galaxies that do not belong to the cluster nor to a filament or a group. 
\par
Here, we adopt an approach similar to that of \cite{Castignani+2022_catalogue}. For the observations, we use their exact definition for cluster and group galaxies, while we redefine the pure field sample by using the same method but considering our definition of filament members.
\par
To identify these same environments for GAEA galaxies, we proceed as follows. 
For each simulated region  (GAEA V1, GAEA V2, and GAEA V3), we define as cluster members those galaxies with a clustercentric distance $<$ 3$R_{200}$ Mpc/h in 3D. 
As mentioned above, we exclude these galaxies when defining filament membership. 
To identify groups, we do not consider the true halo memberships provided by the model as this membership definition would be very different from the observational one, based on a compilation of available observations with different depths. In an attempt to reproduce the observations, 
we select from the GAEA samples V1, V2, and V3 all halos that have at least one galaxy member in our samples. 
We then computed the number of galaxies in each of these halos. 
We defined a group as any gravitationally bounded structure with more than one galaxy $\log [   \rm{M}_{\star}/ \rm{M}_{\sun} ] > 8.3$. Given that this approach to select groups is still different from the observed one, we avoided considering a finer division in groups based on their richness and simply separated isolated galaxies from aggregations of any size. 

Finally, we defined pure field galaxies as those galaxies not belonging to any filaments 
nor to any group or cluster.
\par
We checked if pure field galaxies are actually members of filaments with a density contrast lower than that of the adopted persistence level. However, only 10\% of the pure field galaxies are members of the filaments identified using a 3$\sigma$ persistence level.

\section{Galaxy samples in observations and GAEA} 
\label{sec:finalisingdata}

In the previous sections we have introduced the analysed samples and provided a characterisation of the environments we are going to consider in this work. 
In this section we finalize the galaxy samples we will use, and introduce some definitions useful for the  analysis presented below. 

\subsection{Observational gas mass limit }
\label{sec:gasmass_incomplet_cuts}

\begin{figure}
    \centering
    \includegraphics[width=1\linewidth]{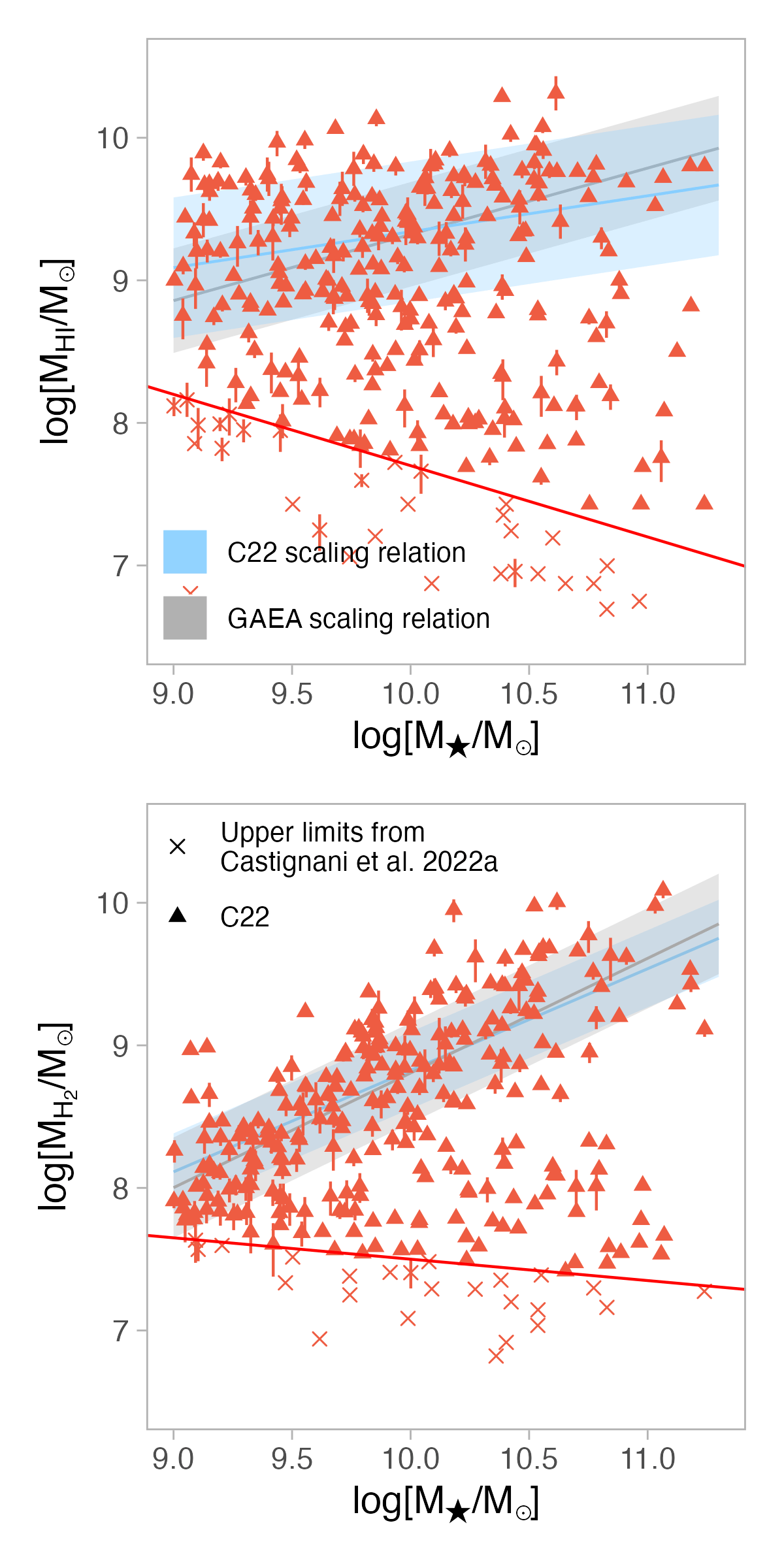}
    \caption{The scaling relation between $\rm{M}_{\ce{HI}}$~(top) and $\rm{M}_{\ce{H2}}$~(bottom) as a function of stellar mass for observations. Red triangles show measurements, and crosses show upper limits. The solid red line shows the adopted level of completeness limits of the observations for \ce{HI}~(Eq.~\ref{eq:limit_h1}) and \ce{H2}~(Eq.~\ref{eq:limit_h2}) masses, respectively. 
    Shaded areas show the scaling relations for C22 (light blue, Eq.~\ref{eq:mhi_mass_rel_obse} in the top panel, Eq.~\ref{eq:mh2_mass_rel_obse} in the bottom panel) and for GAEA (light grey, Eq.~\ref{eq:mhi_mass_rel} in the top panel, Eq.~\ref{eq:mh2_mass_rel} in the bottom panel). 
    }
    \label{fig:hi_h2_cutslvls}
\end{figure}

As mentioned in Sec~\ref{subsec:obs_data}, measurements of atomic and/or molecular hydrogen are not available for all galaxies in the catalogue by \cite{Castignani+2022_catalogue}: some of them have simply not been observed, while for some others only upper limits have been obtained, given their low gas content. 
The limit down to which the gas mass could be obtained for all galaxies depends on many parameters: in terms of fluxes, it depends on the integration time and telescope sensitivity, but in terms of masses, it also depends on full width at half maximum (FWHM) of the detected signal, distance of the source, the observed CO transition, the gas excitation, and aperture correction. 
To perform a meaningful comparison between observational and simulated data, it is important to mimic the observed gas mass completeness of the catalogue, separately for the \ce{HI} and \ce{H2} masses. Given the nature of the observations gathered by \cite{Castignani+2022_gas}, who collected literature data in addition to their own campaigns, it is not possible to properly determine the completeness limits. We, therefore, manually select the  \ce{HI} and \ce{H2} levels above which we consider GAEA and observational samples as complete. Figure~\ref{fig:hi_h2_cutslvls} separately shows the  \ce{HI} and \ce{H2} masses as a function of stellar mass for observations and the thresholds we adopt as completeness limit. The separation was obtained as the line that best separates actual measurements from upper limits:
\begin{equation}
     \log [\rm{M}_{\ce{HI}}/ \rm{M}_{\sun}]  > -0.5 \cdot \log [\rm{M}_{\star}/ \rm{M}_{\sun}]  + 12.7,
    \label{eq:limit_h1}
\end{equation}
\begin{equation}
  \log [\rm{M}_{\ce{H2}}/ \rm{M}_{\sun}]  > -0.15 \cdot \log [\rm{M}_{\star}/ \rm{M}_{\sun}] + 9.0.
     \label{eq:limit_h2}
\end{equation}
We additionally checked that changing this level does not affect the results of this paper.

From now on, in both observations and in the model, we will use only the galaxies with $\rm{M}_{\ce{HI}}$ or $\rm{M}_{\ce{H2}}$ above the separation lines in Fig.~\ref{fig:hi_h2_cutslvls},
and with stellar masses  in the range $ 10^{9} < \rm{M}_{\star}/\rm{M}_{\sun} < 10^{11}$.  We will conduct all the analysis on the  \ce{HI} (\ce{H2}) content considering only the galaxies above the $\rm{M}_{\ce{HI}}$ ($\rm{M}_{\ce{H2}}$) level, regardless of the  \ce{H2} (\ce{HI}) content. When both the \ce{HI} and \ce{H2} will be considered simultaneously, we will use only the galaxies with both $\rm{M}_{\ce{HI}}$ and $\rm{M}_{\ce{H2}}$  above the corresponding levels.

\begin{figure*}
    \centering
    \includegraphics[width=1\linewidth]{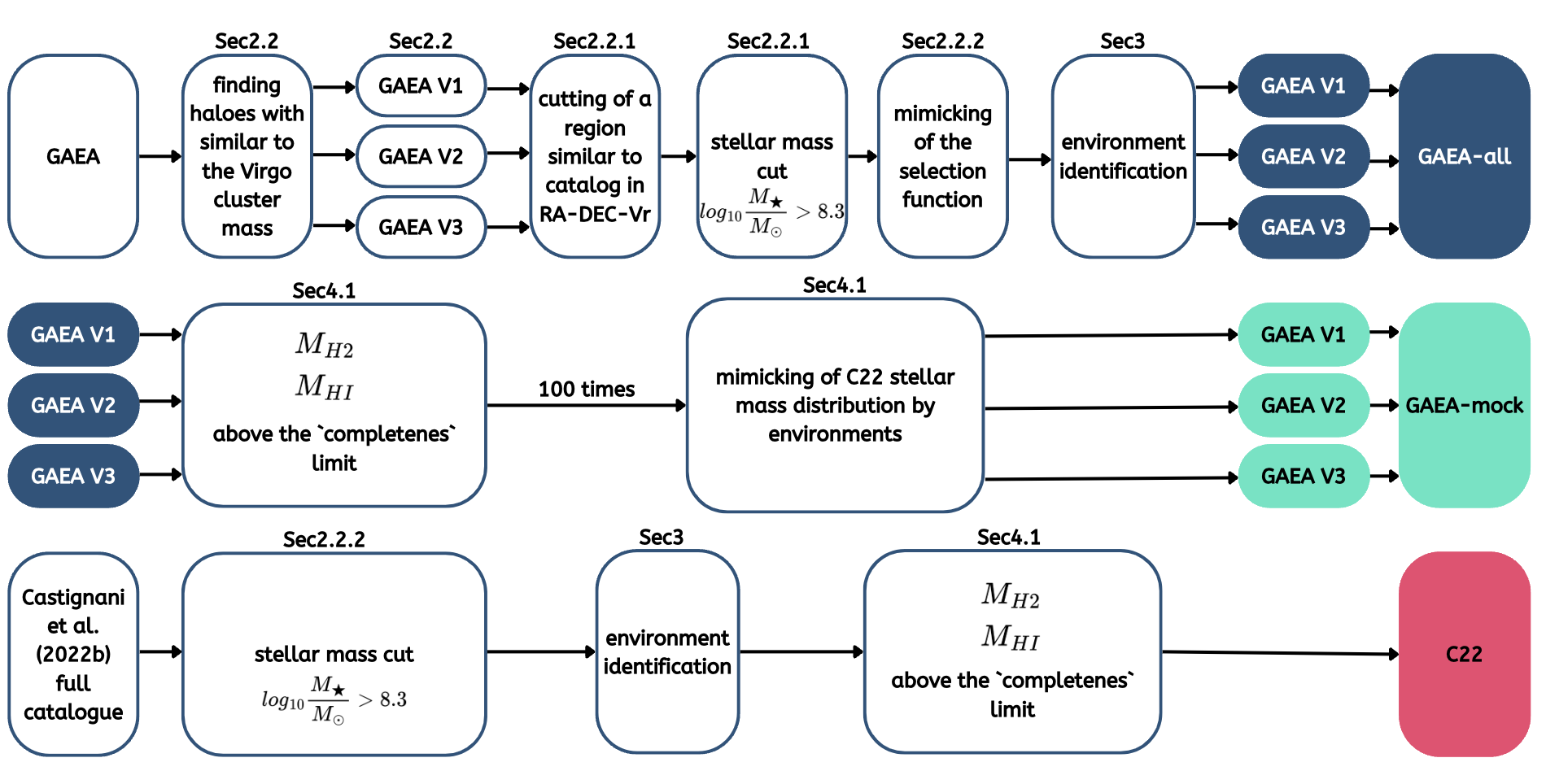}
    \caption{Schematic view of the steps needed to obtain the final sample for both the model (GAEA-all, GAEA-mock) and the observed sample (C22). Each step is described in the main text (Sect 2, 3, 4). }
    \label{fig:dataprep_diagram}
\end{figure*}


Finally, since the stellar mass distribution of the observed galaxies could have an impact on the results, considering each environment separately, we randomly extract from the model samples with the same stellar mass distribution as that of the observed sample. Specifically, we randomly select the same stellar mass distribution 100 times for each environment in the   GAEA V1, GAEA V2, and GAEA V3 cubes. We will call the GAEA sample with this adopted gas limit and stellar mass distribution  `GAEA-mock`.  We will call the observed sample drawn from  \cite{Castignani+2022_catalogue} as explained in the previous section's `C22 sample'.

In addition to the GAEA-mock, we will also consider the GAEA sample relaxing the cut in gas mass, and we will call it `GAEA-all'. This sample also includes only
galaxies with stellar masses  $ \rm{M}_{\star}  > 10^{8.3} \rm{M}_{\sun} $. It will be used to study the impact of observational limits on the results and predict trends in regimes not covered by the current observations. 
We summarize all the performed steps for each of these sets in Fig.~\ref{fig:dataprep_diagram}. 
The number of galaxies in the different samples is given in Table~\ref{tab:number_of_samples_byenv_bymass_hi} and Table~\ref{tab:number_of_samples_byenv_bymass_h2}.


\begin{table*}[]
\caption{Number of galaxies with \ce{HI} measurements above the gas mass completeness limit 
in C22, GAEA-mock, and all in each environment separately.~The model data provides the median number of samples among the three halos under consideration. 
}
\centering
\begin{tabular}{ccccc|c}
&                          & Cluster & Filaments &  Pure field& Total \\
\\
\hline
\\
\multirow{2}{*}{C22} & $ 9 < \log [   \rm{M}_{\star}/ \rm{M}_{\sun} ]  < 10$   & 
26 &  
53      & 
26  & 
102      \\
& $ 10 < \log [\rm{M}_{\star}/ \rm{M}_{\sun}] < 11$ 
& 23& 
41    & 
12 & 
76         \\
\\
\hline
\\
\multirow{2}{*}{GAEA-mock}  &  $ 9 < \log [   \rm{M}_{\star}/ \rm{M}_{\sun} ]  < 10$   &
21 $\pm$ 2        &   
50 $\pm$ 3       &  
22 $\pm$ 2 &
93 $\pm$ 5       \\

& $ 10 < \log [\rm{M}_{\star}/ \rm{M}_{\sun}] < 11$  &
29 $\pm$ 2 &  
46 $\pm$ 3  &    
11 $\pm$ 1    & 
85 $\pm$ 4     \\
\\
\hline
\\
\multirow{2}{*}{GAEA-all}     &  $ \log [   \rm{M}_{\star}/ \rm{M}_{\sun} ]  < 10$   &
799 $\pm$ 72&  
634 $\pm$ 60           &  
699 $\pm$ 35    & 
2076 $\pm$ 12    \\
& $\log [\rm{M}_{\star}/ \rm{M}_{\sun}] > 10$  &  
106 $\pm$ 18     &   
95 $\pm$ 14        &  
83 $\pm$ 15  & 
273 $\pm$ 46       
\end{tabular}

\label{tab:number_of_samples_byenv_bymass_hi}
\end{table*}

\begin{table*}[]
\centering
\caption{Same as Table~\ref{tab:number_of_samples_byenv_bymass_hi} but for \ce{H2}.  GAEA-all is the same as in Table~\ref{tab:number_of_samples_byenv_bymass_hi}. 
}
\begin{tabular}{ccccc|c}
&                          & Cluster & Filaments &  Pure field& Total \\
\\
\hline
\\
\multirow{2}{*}{C22} & $ 9 < \log [   \rm{M}_{\star}/ \rm{M}_{\sun} ]  < 10$   
& 30& 
53      & 
25 &
108      \\

& $ 10  < [\rm{M}_{\star}/ \rm{M}_{\sun}] < 11$  & 
25 &  
41    &  
12 &
78        \\
\\
\hline
\\
\multirow{2}{*}{GAEA-mock}  & $ 9 < \log [   \rm{M}_{\star}/ \rm{M}_{\sun} ]  < 10$  
 & 23 $\pm$ 3        &   
51 $\pm$ 3       &  
    23  $\pm$ 2 &
97 $\pm$ 5      \\

& $ 10 < \log [\rm{M}_{\star}/ \rm{M}_{\sun}] < 11$  &
26 $\pm$ 3 &  
43 $\pm$ 3  
&    
11$\pm$ 2    & 
80 $\pm$ 5     \\
\\     
\end{tabular}

\label{tab:number_of_samples_byenv_bymass_h2}
\end{table*}

\subsection{Definition of gas deficiency} 
\label{sec:deficiency_defs}
A common way to investigate the effect of the environment on the gas content of galaxies is to measure the gas deficiency (e.g. \citealt{Giovanelli+1985}, \citealt{Haynes+1985}, \citealt{Haynes+1986} \citealt{Casoli+1998}, 
\citealt{Chung+2009}, \citealt{Boselli+2014}, \citealt{Hess+2015}, \citealt{Healy+2021}, \citealt{Moretti+2023}), which is defined as the difference between the gas content of a galaxy belonging to a given environment and that of a field galaxy of the same size and morphology.

The exact definition of  \ce{HI} and \ce{H2} deficiency varies from study to study and depends on the specifics of the observational sample (e.g. available information). 
In this work, we base \ce{HI} and \ce{H2} deficiencies on the gas mass versus stellar mass relation. 

In this section, we obtain the \ce{HI} and \ce{H2} scaling relations needed to obtain the \ce{HI} and \ce{H2} deficiency parameters (\ce{HI}$_{\rm{def}}$ and \ce{H2}$_{, \rm{def}}$, respectively) separately in the observations and in the model. The main sequences~(MS) $\rm{M}_{\ce{HI}}-\rm{M}_{\star}$ calculated for the model and data separately helped us not to have to worry about how well the model reproduces the observational MS (although we show below that they are close to each other).

In observations, we separately define scaling relations for \ce{HI} and \ce{H2} using the sample described in Sect. ~\ref{sec:gasmass_incomplet_cuts}. As we 
aim at obtaining a general scaling relation, here we use only star-forming~(specific star formation rate sSFR > $10^{-11}$ year$^{-1}$) field galaxies~($\rm{M}_{halo} < 10^{13} \rm{M}_{\sun}$, with $\rm{M}_{halo}$ mass of group as derived by \citealt{Kourkchi+2017} from the Ks-band luminosity by using M/L ratios), regardless of their filament membership.
By fitting the data using a linear regression method, we obtained the following scaling relations:
 \begin{equation}
        \log [\rm{M}_{\ce{HI}}/ \rm{M}_{\sun}] = 0.25 \cdot \log [\rm{M}_{\star} / \rm{M}_{\sun}] + 6.82 \pm 0.49,
        \label{eq:mhi_mass_rel_obse}
    \end{equation}
  \begin{equation}
        \log [\rm{M}_{\ce{H2}}/ \rm{M}_{\sun}] = 0.8 \cdot \log [\rm{M}_{\star} / \rm{M}_{\sun}] + 0.76 \pm 0.35.
        \label{eq:mh2_mass_rel_obse}
    \end{equation}
Figure~\ref{fig:hi_h2_cutslvls} shows these relations on the top and bottom panels, with a light-blue area marking 1-sigma scatter, respectively.

\par

In GAEA, we defined the scaling relations 
using all the field galaxies in the full cube. As in observations, we considered only star-forming galaxies~(sSFR > $10^{-11}$ year$^{-1}$) that are not part of structures with a halo mass $\rm{M}_{halo} > 10^{13} \rm{M}_{\sun}$ above the gas mass completeness limits~(Fig.~\ref{fig:hi_h2_cutslvls}). 
As before, we fit the data using a linear regression method and obtained the following scaling relations:
     \begin{equation}
        \log [\rm{M}_{\ce{HI}}/ \rm{M}_{\sun}] = 0.47 \cdot \log [\rm{M}_{\star} / \rm{M}_{\sun}] + 4.67 \pm 0.34,
        \label{eq:mhi_mass_rel}
    \end{equation}
    \begin{equation}
        \log [\rm{M}_{\ce{H2}} / \rm{M}_{\sun}] = 0.71 \cdot \log [\rm{M}_{\star}/\rm{M}_{\sun}] + 1.7 \pm 0.26.
        \label{eq:mh2_mass_rel}
    \end{equation}
These relations are also shown in Fig.~\ref{fig:hi_h2_cutslvls} by light-grey areas and are in excellent agreement with the observational determination.

We were then in the position of defining the  \ce{HI} and \ce{H2} deficiencies as the logarithmic difference between the expected~(for a given mass) and measured \ce{HI} and \ce{H2} mass, respectively:
    \begin{equation}
        \ce{HI}_{\rm{def}}  = 
        \log[\rm{M}_{\ce{HI}}^{ \rm{EXP}}/ \rm{M}_{\sun}] -  \log[\rm{M}_{\ce{HI}}^{ \rm{MES}}/ \rm{M}_{\sun}],
        \label{eq:mhi_def} 
    \end{equation}

   \begin{equation}
        \ce{H}_{2,\rm{def}}  = 
        \log[\rm{M}_{\ce{H2}}^{ \rm{EXP}}/ \rm{M}_{\sun}] -  \log[\rm{M}_{\ce{H2}}^{ \rm{MES}}/ \rm{M}_{\sun}],
        \label{eq:mh2_def} 
    \end{equation}
with $\log[\rm{M}_{\ce{HI}}^{ \rm{EXP}}/ \rm{M}_{\sun}]$ obtained from Eq.\ref{eq:mhi_mass_rel_obse} for observations and  Eq.\ref{eq:mhi_mass_rel} for GAEA, $\log[\rm{M}_{\ce{H2}}^{ \rm{EXP}}/ \rm{M}_{\sun}]$ obtained from Eq.\ref{eq:mh2_mass_rel_obse} for observations, and  Eq.\ref{eq:mh2_mass_rel} for GAEA. 
\par 
In this work, we consider a galaxy as \ce{HI}~(\ce{H2}) deficient when $\ce{HI}_{\rm{def}} > 0.5~( \ce{H}_{2,\rm{def}} > 0.5)$, and we consider  \ce{HI}~(\ce{H2}) normal if $\ce{HI}_{\rm{def}} \le 0.5~( \ce{H}_{2,\rm{def}} \le 0.5)$.
\par 
We note that \cite{Castignani+2022_gas} adopted a different definition of deficiency,  
which involves only field galaxies within the same morphological type and optical sizes. Here, we do not adopt their approach to be consistent in definitions between the observations and the model
However,  we compare the deficiency values used by  \cite{Castignani+2022_gas} and ours, finding a good correlation (see Appendix~\ref{app:defs_comparison}). A more thorough  discussion on the different ways to define the expected $\rm{M}_{\ce{HI}}$ or $\rm{M}_{\ce{H2}}$  to estimate deficiencies can be found in 
~\cite{Li+2020_hidef_reciepts} or \cite{Cortese+2021} and is beyond the scope of this work.

\section{Results: \ce{HI} and \ce{H2} content }
\label{sec:results}

In this section we characterize the gas content of galaxies in cluster, filaments and pure field using both the observed data and  the model predictions.  We first consider the atomic hydrogen content and investigate how galaxies are distributed on the $ \rm{M}_{\ce{HI}}-\rm{M}_{\star}$ plane and discuss how the \ce{HI}-deficiency distributions depend on the position of galaxies within the cosmic web. Next, we repeat the same analysis for the molecular hydrogen \ce{H2} content. Finally, we combine the two measurements and contrast \ce{HI} and \ce{H2} deficiency and look for correlations.

\subsection{Atomic hydrogen \ce{HI} content}
\label{sec:redults_hi}

In this section we use only galaxies with  $\rm{M}_{\ce{HI}}$ above the limit given in Eq.~\ref{eq:limit_h1}, both  for C22 and GAEA-mock samples, regardless of their \ce{H2}-content. 

\begin{figure*}
    \centering
    \includegraphics[width=1\linewidth]{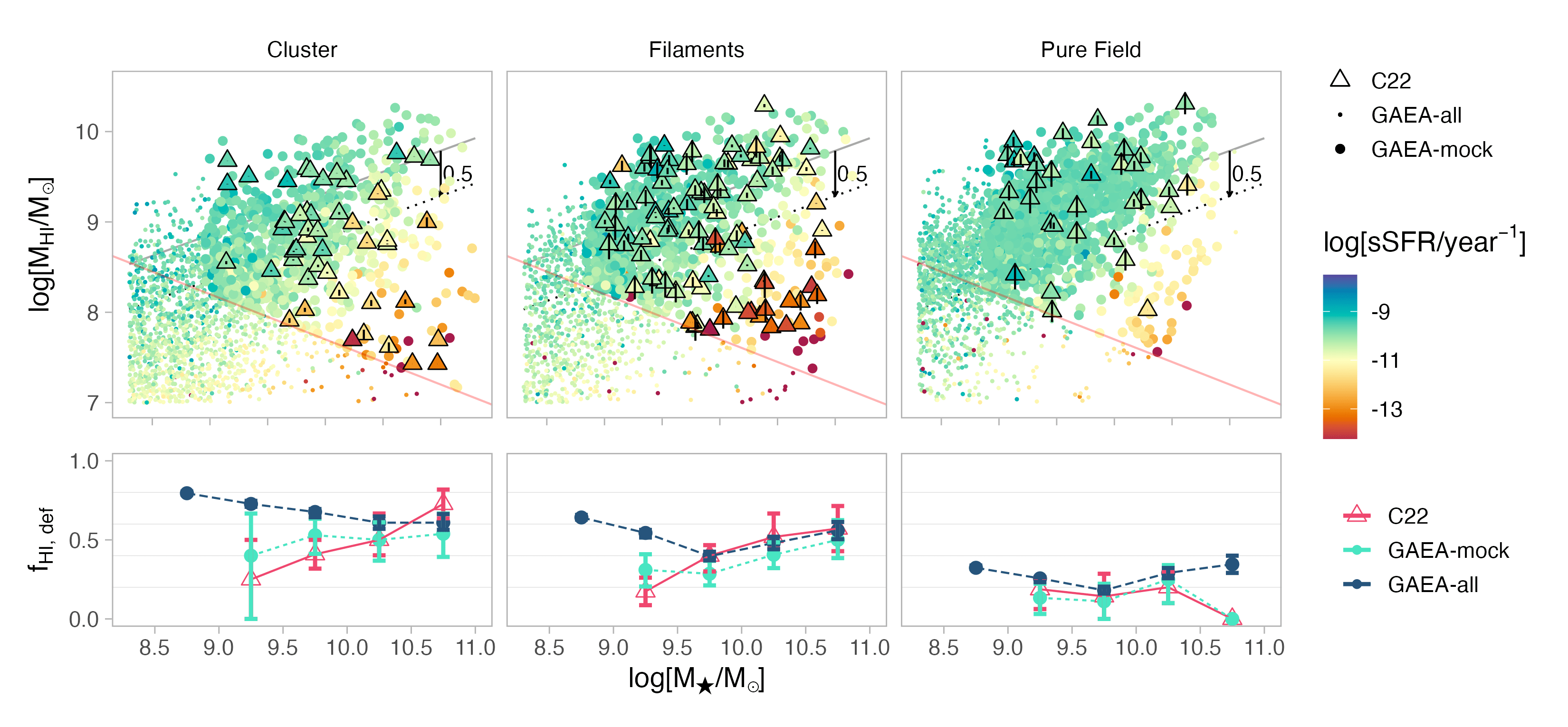}
    \caption{The amount of \ce{HI}-content in galaxies in clusters, filaments, and pure field.
    Top: $ \rm{M}_{\ce{HI}}$ as a function of stellar mass  in different environments, indicated on top of each panel. The GAEA model data are shown with circles: small circles represent the GAEA-all set, and big circles represent one of the 300 realisations of the GAEA-mock sample. Big triangles show the C22 data. In all the samples, each point is coloured by sSFR. In each panel, the solid grey line shows the  $ \rm{M}_{\ce{HI}}-\rm{M}_{\star}$ scaling relation (Eq.~\ref{eq:mhi_mass_rel}), the dotted line shows the 0.5 dex indent to highlight \ce{HI}-deficiency zone.
    The faint red line represents \ce{HI}-mass completeness limit from Eq.~\ref{eq:limit_h1}. 
    Bottom: Fractions of \ce{HI}-deficient galaxies~(see Sect.~\ref{sec:deficiency_defs}) with 1$\sigma$ confidence interval as a function of stellar mass.}
    \label{fig:himass_mass_bymassenvs}
\end{figure*}

The top panel of Fig.~\ref{fig:himass_mass_bymassenvs} shows the relation between $\rm{M}_{\ce{HI}}$ and stellar mass $\rm{M}_{\star}$ for galaxies in different environments for C22, GAEA-mock, and GAEA-all.
In agreement with the vast literature (e.g. \citealt{Catinella+2010, Parkash+2018}), we recover a positive correlation between $\rm{M}_{\ce{HI}}$ and $\rm{M}_{\star}$. Overall, the C22 and GAEA-mock galaxy samples, which by construction are directly comparable, occupy the same region of the plane. Furthermore, while most of the galaxies are concentrated around the scaling relation defined by Eq.~\ref{eq:mhi_mass_rel}, a non-negligible population deviates from it, having a lower $\rm{M}_{\ce{HI}}$  than expected, given their stellar mass. The fraction of cluster galaxies with reduced $\rm{M}_{\ce{HI}}$ is comparable between C22 and GAEA-mock samples and is 49$\pm$7\% and 51$\pm$6\%, respectively. Similarly, also  the relative abundance of pure field galaxies with low levels of $\rm{M}_{\ce{HI}}$ are compatible: in both cases they are 17$\pm$5\%. 
Filament galaxies have intermediate position in terms of reduced amounts of atomic hydrogen in 43$\pm$5\% cases for C22 and 41$\pm$5\% for GAEA-mock. 

Our results are qualitatively consistent with many previous works that  found an increased proportion of galaxies with reduced $\rm{M}_{\ce{HI}}$ in clusters compared to  field galaxies of similar mass \citep{Haynes+1985,Boselli+2006} and that filament galaxies occupy an intermediate position between cluster and  field galaxies~(e.g. \citealt{Blue_Bird+2020, Castignani+2022_gas}). In addition, we note that the model does reproduce observational data for galaxies in filaments, although the model does not include a special treatment for filaments.
\par
When considering GAEA-all, we find that the fraction of low \ce{HI}-content galaxies decreases from clusters~(74$\pm$1\%) to filaments~(57$\pm$2\%) to pure field~(32$\pm$1\%). However, the absolute numbers are higher than for C22 and GAEA-mock, which is obviously due to the adopted gas mass limit in Sect.~\ref{sec:gasmass_incomplet_cuts}.

The bottom panel of Fig.~\ref{fig:himass_mass_bymassenvs} shows  how the fraction of \ce{HI}-deficient~
galaxies depends on stellar mass, where median fractions of $\rm{HI}_{def}$ with $1\sigma$ confidence interval for each mass bin are reported. C22 and GAEA-mock provide a consistent picture: overall, where there is enough statistics, the \ce{HI}-deficient galaxy fraction increases with increasing stellar mass, except in the observed Virgo cluster where it is consistent with being flat across the considered mass range. The GAEA-all sample, which allowed us to get rid of some observational biases, shows instead different trends. In the cluster and in filaments,  the fraction of \ce{HI}-deficient galaxies decreases with increasing stellar mass. This result is due to the fact that in GAEA low-mass galaxies are more affected by ram-pressure stripping forces because of their low restoring force~\cite{GunnGott1972}. As a consequence, they have a higher probability of being \ce{HI}-deficient (see also \citealt{Xie+2020}). 
\par
The points in the top panel of Fig.~\ref{fig:himass_mass_bymassenvs} are coloured by the galaxy  sSFR. In general, in both the GAEA-mock and C22 samples
galaxies with relatively low $\rm{M}_{\ce{HI}}$  are also characterised by  low sSFR values. This is particularly true for high-mass galaxies.  There are though some exceptions, with galaxies with normal \ce{HI} gas content having already low sSFR values (in agreement with e.g. \citealt{Zhang+2019}) and, vice-versa, galaxies with reduced  $\rm{M}_{\ce{HI}}$ but with high sSFR values, indicative of non-negligible, ongoing star formation. However, this is not unexpected: star formation is found to be more strongly correlated with the surface density of molecular hydrogen than with atomic hydrogen~\citep{Leroy+2008}.


\begin{figure*}
    \centering
     \includegraphics[width=1\linewidth]{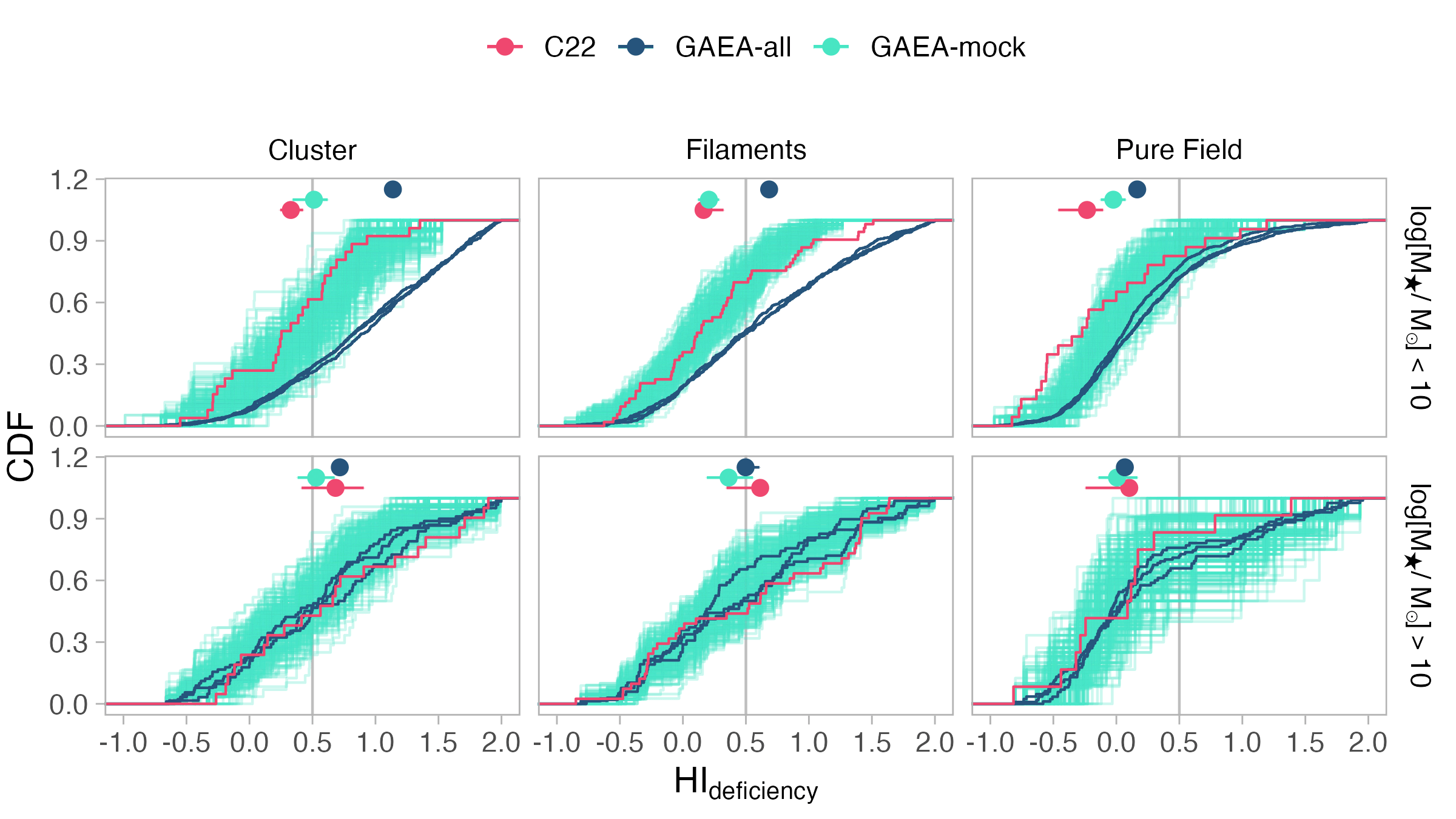}
    \caption{\ce{HI}-deficiency cumulative distribution function for GAEA-all~(three lines for GAEA-all V1, GAEA-all V2 and GAEA-all V3); GAEA-mock~(300 lines); and C22 split into the different environments and two mass bins~(low-mass $\log_{10} [{\rm M}_{\star} / \rm{M}_{\sun} ] < 10$ and massive galaxies $ \log_{10} [{\rm M}_{\star} / \rm{M}_{\sun} ]> 10$). In each plot, the median values with a 1$\sigma$ confidential interval are reported. The grey vertical line shows the 0.5 \ce{HI}-deficiency level as a level adopted to consider a galaxy as \ce{HI}-deficient. 
    }
    \label{fig:mhidef_bymass_byenv}
\end{figure*}

The results presented above rely on the adopted separation between \ce{HI}-normal and \ce{HI}-deficient galaxies. To obtain more general results, in Fig.~\ref{fig:mhidef_bymass_byenv} we consider the distribution of the \ce{HI}-deficiency, and investigate how the whole population of galaxies behaves in the different environments. To account for the dependence on stellar mass in Fig.~\ref{fig:himass_mass_bymassenvs}, we then consider two mass bins: low-mass~($ \log_{10} [{\rm M}_{\star} / \rm{M}_{\sun} ] < 10$) and massive~($ \log_{10} [{\rm M}_{\star} / \rm{M}_{\sun} ] > 10$) galaxies.
\par
Overall, the cumulative distribution function of the \ce{HI}-deficiency  in the GAEA-mock is compatible to that obtained for C22, in  each environment and in both mass bins.   
Considering the low-mass galaxies, C22 and GAEA-mock retrieve consistent trends:  43$\pm$8\% and 52$\pm$10\%  of the cluster population have a \ce{HI}-deficiency parameter $>0.5$ dex. Moving to filaments and pure field, the median values of the distributions shift to lower values, indicating galaxies are most likely \ce{HI}-normal~(only 29$\pm$6\%~(13$\pm$7\%) and 30$\pm$6\%~(17$\pm$9\%) of low-mass filaments~(pure field) galaxies are \ce{HI}-deficient for C22 and GAEA-mock, respectively). We additionally confirm the correspondence between C22 and GAEA-mock low-mass galaxies in terms of \ce{HI}- deficiency running the KS test pairwise on C22 and each of the 300 realisations of the GAEA-mock sample. Considering each environment separately, we find that distributions are indistinguishable (p-v$>0.05$) in at least 90\% of the cases for cluster and filaments samples, and only in 77\% for pure field. 
\par
The comparison between  GAEA-all and C22/GAEA-mock in Fig.~\ref{fig:mhidef_bymass_byenv} shows how the adopted gas mass  limit affects the completeness of the low-mass galaxy population. 
GAEA-all predicts a much more substantial  fraction of \ce{HI}-deficient low-mass galaxies in cluster~(76$\pm$1\%), in filaments~(61$\pm$2\%), and in pure field~(30$\pm$1\%) than observed.
\par
The case of massive galaxies  is similar to the low-mass one: the fraction of \ce{HI}-deficient galaxies is the greatest for cluster members~(C22 shows 61$\pm$8\%, GAEA-mock have 52$\pm$9\% \ce{HI}-deficient massive galaxies) and declines to filaments~(56$\pm$6\% for C22 and 44$\pm$7\% for GAEA-mock) and pure field~(16$\pm$8\% for C22 and 25$\pm$11\% for GAEA-mock) although clusters and filaments fractions are compatible within errors. 
Figure~\ref{fig:mhidef_bymass_byenv} shows a correspondence between the cumulative distribution functions of \ce{HI}-deficiency of massive galaxies in C22 and GAEA-mock within each environment. We confirm this result with the KS test, which reports that distributions are indistinguishable ~(p-v$>0.05$) in 99\%, 87\%, and 97\% of the cases for cluster, filaments, and pure field galaxies, respectively. Due to the adopted gas mass completeness limit, we do not observe any significant difference between GAEA-all and C22/GAEA-mock for massive galaxies.  Overall, GAEA-all predicts 62$\pm$6\%, 54$\pm$5\%, and 32$\pm$4\% of \ce{HI}-deficient massive galaxies in clusters, filaments, and pure fields, respectively.
\par
To summarize, we find an excellent agreement between the C22 and GAEA-mock samples and also detect for all three sets a decrease in the proportion of \ce{HI}-deficient galaxies from clusters to filaments and to the pure field. However, we do not find a significant difference between massive and low-mass galaxies in C22/GAEA-mock in cluster and pure field
, although GAEA-all predicts that the proportion of \ce{HI}-deficient low-mass galaxies is higher than the proportion of massive ones within cluster and filaments.


\subsection{ Molecular hydrogen \ce{H2}}
\label{sec:results_h2}

Next, we focus on the \ce{H2} content of galaxies. For C22 and GAEA-mock, we used only galaxies with  $\rm{M}_{\ce{H2}}$ above the limit Eq.~\ref{eq:limit_h2},\footnote{Varying the adopted completeness level does not impact the results.} regardless of their \ce{HI}-content. 
\begin{figure*}
    \centering
    \includegraphics[width=1\linewidth]{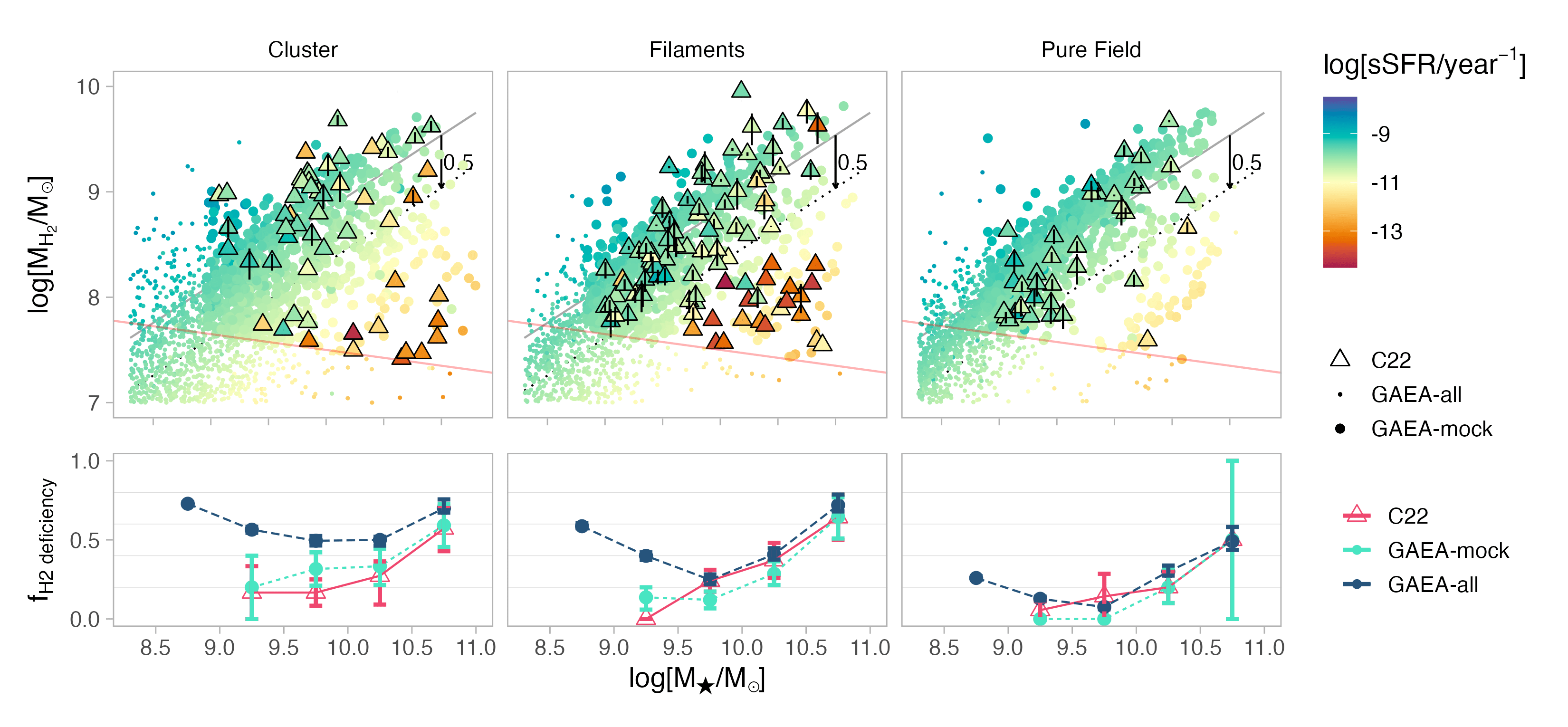}
    \caption{The amount of \ce{H2}-content in galaxies in clusters, filaments, and pure field. Top: $ \rm{M}_{\ce{H2}}$ as a function of stellar mass in the different environments. Panels, colours, and symbols are as in Fig. \ref{fig:himass_mass_bymassenvs}. 
    Bottom: Fractions of \ce{H2}-deficient galaxies in each environment by mass bins. Panels, colours, and symbols are as in Fig. \ref{fig:himass_mass_bymassenvs}.  
    }
    \label{fig:h2mass_mass_bymassenvs}
\end{figure*}


\begin{figure*}
    \centering
    \includegraphics[width=1\linewidth]{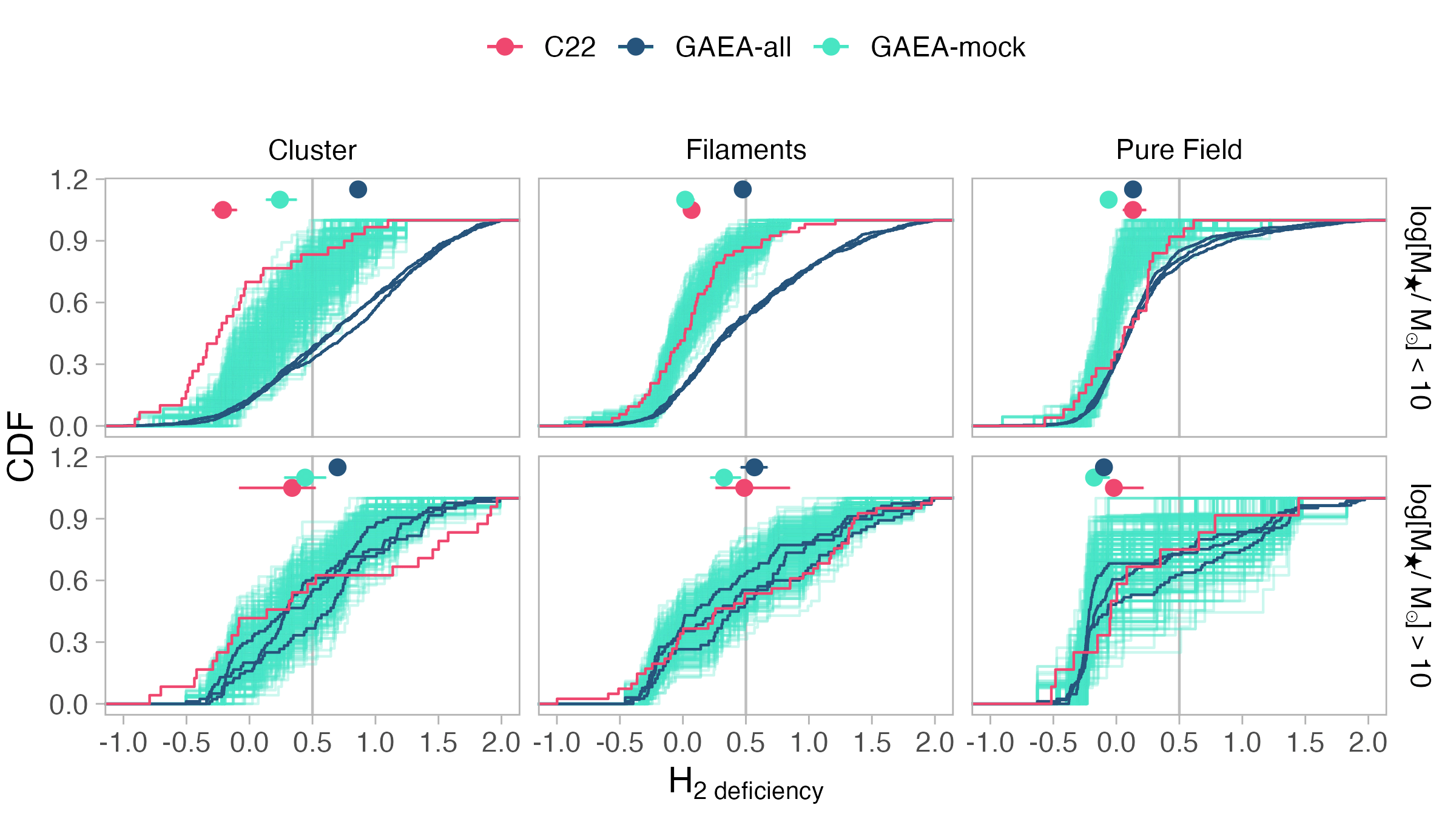}
    \caption{\ce{H2}-deficiency distributions~(CDF) and median values with 1$\sigma$ confidential interval. The caption is the same as Fig.~\ref{fig:mhidef_bymass_byenv}. The number of samples is presented in Table.~\ref{tab:number_of_samples_byenv_bymass_h2}. }
    \label{fig:mh2def_bymass_byenv}
\end{figure*}

The top panel of Fig.~\ref{fig:h2mass_mass_bymassenvs}  shows the mass of molecular hydrogen $\rm{M}_{\ce{H2}}$ as a function of stellar mass for galaxies in different environments for GAEA and C22. 
As in the case of $\rm{M}_{\ce{HI}}$,  we recover a correlation between $\rm{M}_{\ce{H2}}$-$\rm{M}_{\star}$for all the considered environments and a good agreement between C22 and GAEA-mock samples. A significant \ce{H2}-deficient population in cluster and filaments emerges. In GAEA-mock, the fraction of \ce{H2}-deficient galaxies (\ce{H2}$_{,\rm def}$ > 0.5) decreases from cluster (38$\pm$7\%) to filaments  (26$\pm$4\%) and pure field  (14$\pm$4\%) galaxies. Similarly, C22 shows the fraction of \ce{H2}-deficient galaxies is 29$\pm$7\% in the cluster,  26$\pm$4\% in the filaments, and 13$\pm$5\% in the pure field. The bottom panel of Fig.~\ref{fig:h2mass_mass_bymassenvs} shows the fraction of \ce{H2}-deficient galaxies in different environments and mass bins. GAEA-mock and C22  provide consistent results within errors for each of the environments: the fraction of \ce{H2}-deficient galaxies rapidly increases with increasing stellar mass. At the same time, GAEA-all predicts an  U-like shape in the fraction of \ce{H2}-deficient galaxies, where galaxies with  $\log_{10} [{\rm M}_{\star} / \rm{M}_{\sun} ] < 9$ or $\log_{10} [{\rm M}_{\star} / \rm{M}_{\sun} ] > 10.5$ have a higher probability of being \ce{H2}-deficient. 
\par

The  correspondence between the GAEA-mock and C22 for the molecular hydrogen allowed us to make predictions according to GAEA-all for the fraction of \ce{H2}-deficient galaxies without observational biases: 65$\pm$2\% of cluster galaxies 51$\pm$2\% of filament galaxies and 25$\pm$1\% of the pure filament galaxies are  \ce{H2}-deficient.  We note that the fractions of  \ce{H2}-deficient galaxies are lower than those of \ce{HI}-deficient galaxies, in  each environment separately~(by $\approx$10\% for cluster and filaments and by $\approx$5\% for pure field). This is most likely due to the fact that by design  GAEA  includes the removal of \ce{HI} ahead of \ce{H2}, to match observational results~(see \citealt{Boselli+2014} or \citealt{Cortese+2021} for a review).
\par
Points in Fig.~\ref{fig:h2mass_mass_bymassenvs}~(left panel) are colour-coded by  sSFR values.  As expected, \ce{H2}-deficient galaxies are also quiescent, in all environments~\citep{Leroy+2008}. 
Nevertheless, in C22  20\% of cluster galaxies with normal \ce{H2} content~
are quiescent, while GAEA does not predict the existence of this population.


\begin{figure*}
    \centering
    \includegraphics[width=1\linewidth]{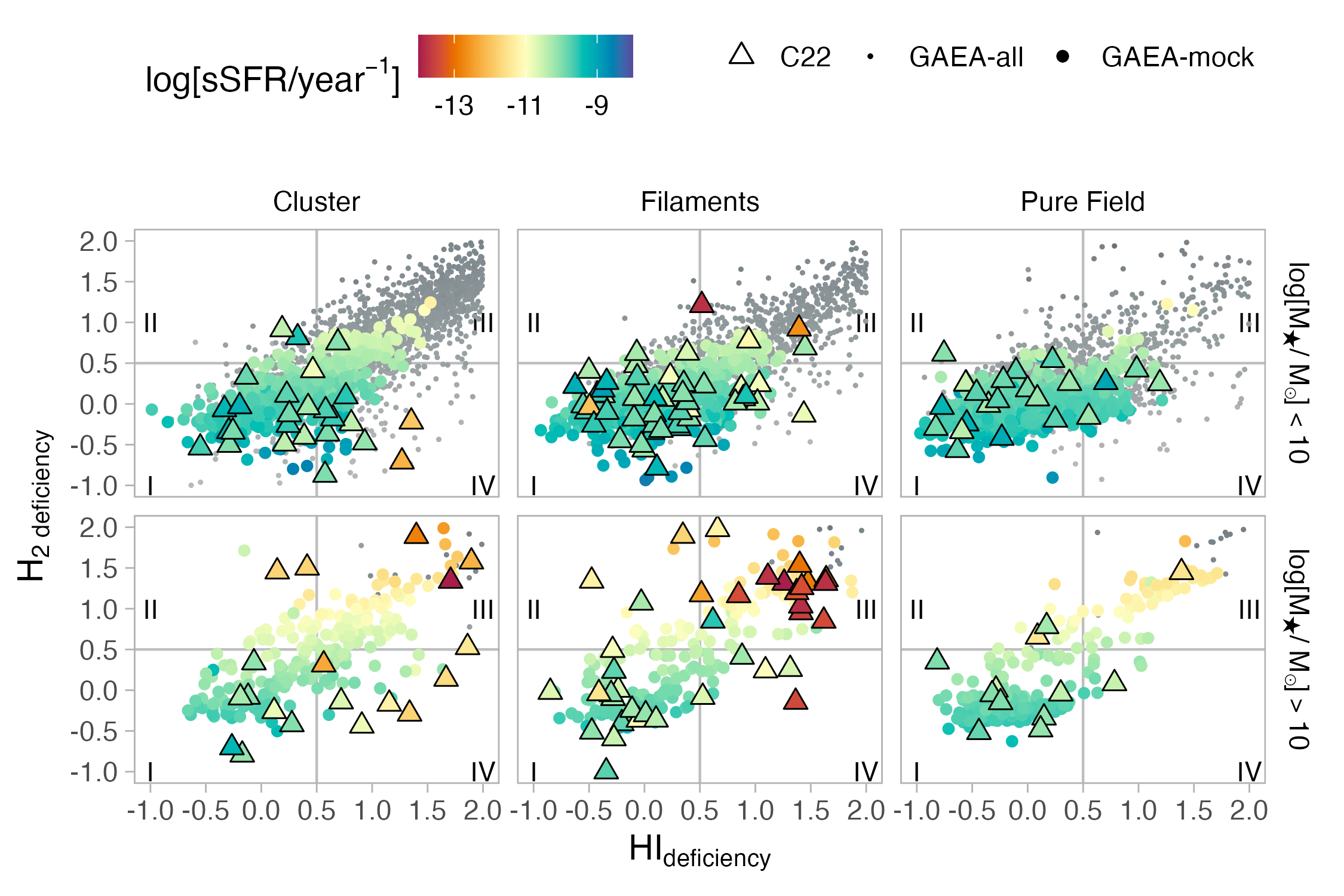}
    \caption{\ce{H2}$_{, \rm{deficiency}}$-\ce{HI}$_{\rm{deficiency}}$ relations for low-mass  (top) and massive (bottom) galaxies 
    in clusters (left), filaments (middle), and the pure field (right). GAEA-mock data is represented by big circles, GAEA-all data by small circles, and C22 data by triangles. Each point of the GAEA-mock/C22 is coloured by sSFR. The vertical and horizontal lines show 0.5 dex deficiency levels used to separate gas normal from gas deficient galaxies. }
    \label{fig:mhidef_mh2def_bymass_byenv}
\end{figure*}

Figure~\ref{fig:mh2def_bymass_byenv} shows the cumulative distribution function of \ce{H2}-deficiency for galaxies in various environments in two mass bins for GAEA-all, GAEA-mock, and C22. %
According to GAEA-mock,  clusters have a higher fraction of  low-mass  ($\log_{10} [{\rm M}_{\star} / \rm{M}_{\sun} ] < 10$) \ce{H2}-deficient  galaxies than  filaments and pure field: 28$\pm$9\%, 12$\pm$4\% and 4$\pm$4\%, respectively. This is in broad agreement with the C22 sample where fractions are 16$\pm$6\%, 13$\pm$4\%, and 8$\pm$4\% for low-mass galaxies in the same environments. Overall, these values are significantly lower than those obtained for the \ce{HI}-deficiency, both in observations and in the model, indicating that  \ce{HI} must be removed more rapidly than \ce{H2}, in all environments and regardless of the mechanisms affecting the gas content.
\par
To further check the correspondence between the GAEA-mock and C22 for low-mass galaxies, we run the KS test between the different  distributions. The KS test reveals that GAEA-mock does not reproduce C22 \ce{H2}-deficiency properly for cluster and pure field: only 6\% and 30\%  of GAEA-mock samples have indistinguishable distributions from observed ones. In contrast, when comparing distributions in filaments, we retrieve no difference  between the model and C22 in 98\% of the GAEA-mock realisations.  
\par
Considering massive galaxies ($\log_{10} [{\rm M}_{\star} / \rm{M}_{\sun} ] > 10$), we obtain  the following fractions of \ce{H2}-deficient galaxies  for C22 and GAEA-mock respectively: 44$\pm$12\% and 44$\pm$9\% for cluster galaxies ; 46$\pm$9\% and 41$\pm$8\% for filament galaxies, and 25$\pm$13\% and 22$\pm$12\% for pure field galaxies. Thus, massive galaxies have a similar fraction of \ce{H2}-deficiency in all the environments. Also, C22~(but not GAEA-mock) shows the same values for low-mass ones in filaments and the cluster. 
\par
The KS test between C22 and GAEA-mock for massive galaxies shows a good correspondence for \ce{H2}-deficiency distribution in 69\%, 86\%, and 88\% for the cluster, filaments, and pure field galaxies. Considering the GAEA-all sets, we obtain a similar trend of decreasing fraction of \ce{H2}-deficient galaxies from the cluster to filaments and to the pure field for low-mass~(67$\pm$1\%, 52$\pm$1\% and 23$\pm$1\%, respectively)  and massive galaxies~(56$\pm$4\%, 52$\pm$4\% and 38$\pm$6\%, respectively). 
\par
Taking into account observational biases, we conclude that the GAEA model is reproducing how the \ce{H2}-deficiency depends on the environment, especially for massive galaxies.

\subsection{\ce{HI}- vs \ce{H2}-deficiency}

Having established that similar \ce{H2}-deficiency and \ce{HI}-deficiency trends are found in   GAEA-mock and C22,  we then combine the \ce{HI}-deficiency and \ce{H2}-deficiency measurements.  Figure~\ref{fig:mhidef_mh2def_bymass_byenv} shows the  {\ce{H2}$_{, \rm{def}}$-\ce{HI}$_{\rm{def}}$  relation for galaxies in different environments and within two stellar mass bins, considering only galaxies with both \ce{HI} and \ce{H2} masses above the corresponding thresholds~\footnote{For C22 and GAEA-mock}~(Eq.~\ref{eq:limit_h1} and \ref{eq:limit_h2}).  
Each panel shows the separation between \ce{HI}-deficient/\ce{H2}-deficient and \ce{HI}-normal/\ce{H2}-normal regions,
so each panel is separated into four quadrants:  I~(\ce{HI}- and \ce{H2}-normal galaxies), II~(\ce{HI}-normal and \ce{H2}-deficient galaxies), III~(\ce{HI}- and \ce{H2}-deficient galaxies), IV~(\ce{HI}-deficient and \ce{H2}-normal galaxies). 
\par
For all samples, Fig.~\ref{fig:mhidef_mh2def_bymass_byenv}  shows a clear correlation between \ce{H}$_{2, \rm{def}}$ and \ce{HI}$_{\rm{def}}$ for each environment and mass bins, in agreement with other works (e.g. \citealt{Zabel+2022, Moretti+2023}). The proportion of galaxies with deficiency only in one gas phase according to GAEA-all does not exceed 20\% across all environments and mass bins~(17$\pm$1\% for low-mass galaxies and 15$\pm$3\% for massive ones). The GAEA-all sample, plotted in the background, highlights the already discussed observational biases emerging for low-mass and \ce{HI}- and \ce{H2}-deficient~(III quadrants) galaxies: GAEA-all predicts a population of low-mass \ce{HI}- and \ce{H2}-deficient galaxies, which are under the gas mass completeness limits. Comparing C22 and GAEA-mock, we find that low-mass galaxies are mostly located in the first quadrant. Moreover, low-mass galaxies are star-forming regardless of the surroundings, either for C22 or GAEA-mock: more than 90\% if galaxies in  each environment separately have sSFR > $10^{-11}$ year$^{-1}$.
\par
The top panels of Fig.~\ref{fig:mhidef_mh2def_bymass_byenv} also show how \ce{HI} and \ce{H2}-deficiency for low-mass galaxies  depends on the environment according to GAEA-all: the fractions of galaxies inside III quadrant decline from cluster to filaments and pure field: 62$\pm$2\%, 45$\pm$1\% and 16$\pm$1\%, respectively. The third quadrant contains mostly galaxies with low sSFR~(70\% of III quadrant galaxies are quiescent with sSFR < $10^{-11}$ year$^{-1}$), which is expected since the absence of cold gas is inevitably connected to suppressed star formation in galaxies~\citep{Leroy+2008, Boselli+2014}.
\par

\begin{figure}
    \centering
    \includegraphics[width=1\linewidth]{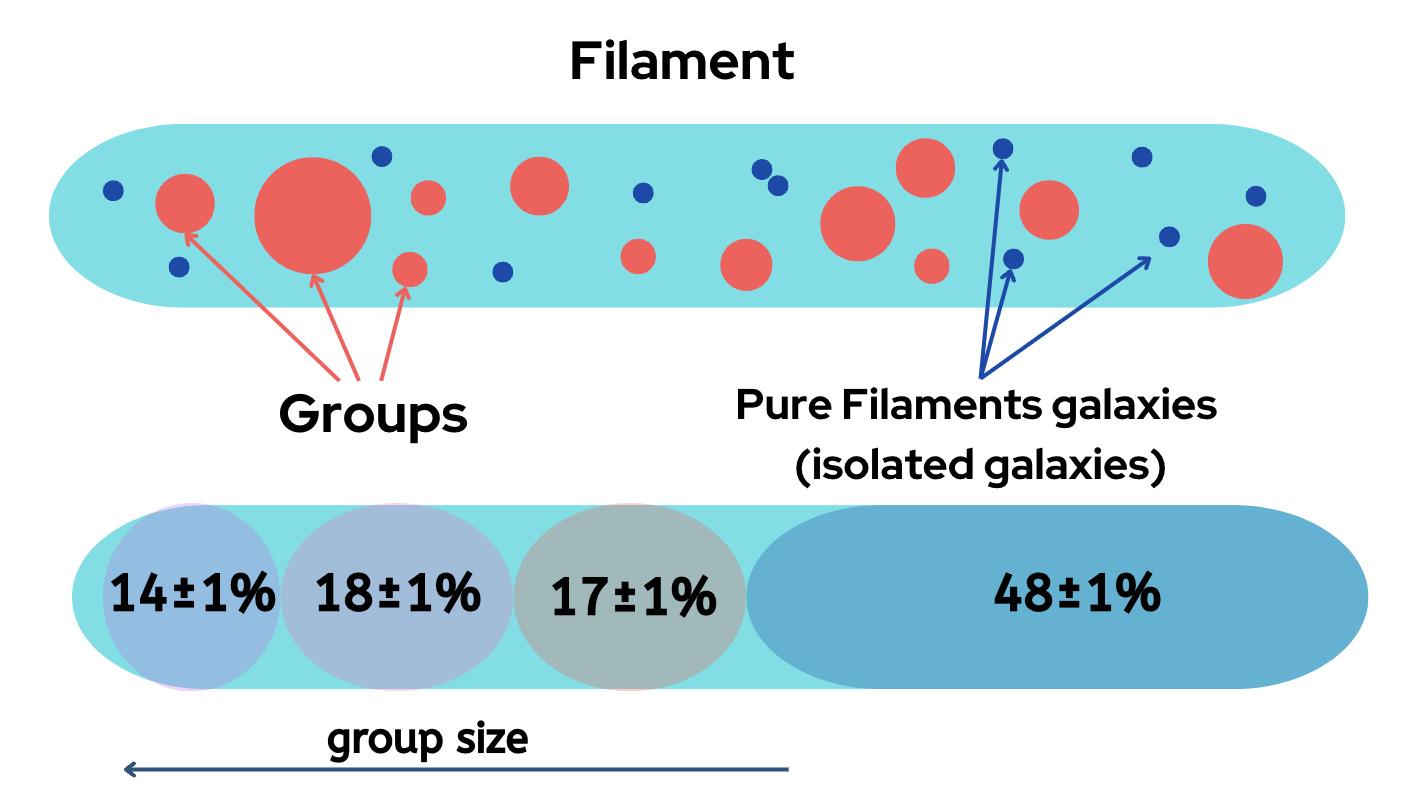}
    \caption{Schematic representation of filament composition by galaxies in groups of different sizes and pure filament galaxies (i.e. alone in their halo; top panel). The proportion of galaxies in groups with 15 < $\rm N_{\rm mem}$ < 50, 5 < $\rm N_{\rm mem}$ < 15, and 1 < $\rm N_{\rm mem}$ < 5 members and pure filament in total filaments population, respectively~(bottom panel), taking into account the selection function.}
    \label{fig:scheme}
\end{figure}

Among massive galaxies,  C22 and GAEA-mock mostly occupy the first and the third quadrants, as well as GAEA-all. 
However, GAEA-all predicts a smaller difference in fractions of \ce{HI} and \ce{H2}-deficient galaxies between environments for massive galaxies (as was discussed above) than for low-mass systems: 48$\pm$5\%, 49$\pm$1\% and 26$\pm$4 for the cluster, filaments, and pure field. 
GAEA-mock follows this trend with 33$\pm$9\% of the cluster, 33$\pm$8\% filaments, and 17$\pm$9\% pure field galaxies inside the III quadrant. C22 sample shows 24$\pm$9\%, 39$\pm$8\%, and 17$\pm$9\% percents of  \ce{HI}- and \ce{H2}-deficient in cluster, filaments, and field. We stress that here we are considering only galaxies above both \ce{HI} and \ce{H2} thresholds and we are therefore using a subsample of that used in the previous section. This is the reason why results seems at odds with what previously shown.
\par
In Sect.~\ref{sec:results_h2} we discussed the presence of \ce{H2}-normal galaxies but without ongoing star formation. Figure~\ref{fig:mhidef_mh2def_bymass_byenv}  reveals that those galaxies are strongly \ce{HI}-deficient galaxies. It is surprising since we do not expect that \ce{HI}-deficiency is sufficient to prevent star formation in galaxies with normal \ce{H2}. 

\par
To summarize the entire section, when considering as much as possible the several observational biases, and the inability to accurately reproduce the selection function, the model reproduces the \ce{HI} content in galaxies for each of the three considered environments when separated into low-mass and massive galaxies. The replication by the model of \ce{H2} content in filaments and in the pure field as well is good, but for cluster galaxies, we obtained uncertainty in \ce{H2}-deficiency, which is apparently related to observational biases. 
Finally, we show that the difference in the proportion of \ce{HI} and/or \ce{H2}-deficiency of low-mass galaxies across all the environments is more pronounced than for massive galaxies. So, massive galaxies are less sensitive to environments than low-mass galaxies either for \ce{HI}- and \ce{H2}-deficiency. In addition, both the model and observations show that the difference in \ce{HI}-deficiency between environments is more pronounced  than \ce{H2}-deficiency, suggesting that \ce{HI} is more sensitive to the environment, in agreement with previous works  ~\citep{Boselli+2014, Loni+2021}.



\section{Discussion}
\label{sec:discussion}

In the previous section we have shown that results obtained using  the GAEA model and observations are in broad agreement; we discuss the influence of filaments on the evolution of galaxies using only the GAEA-all dataset, which is not affected by observational biases. We will examine how filaments influence galaxy evolution in terms of assembly history~(at fixed halo mass, the assembly of dark matter haloes correlates with the large-scale environment~\citep{Gao+2005}, which in turn is imprinted on the assembly of galaxies~\citep{Croton+2007}).
We again emphasize that the GAEA model does not include any specific treatment for galaxies in filaments other than assembly bias. This means that there is no  predetermined dependence of galaxy properties on the distance to the axis of the filaments,  galaxy-filament interaction is not considered and there are no specific modes of accretion of cold gas to galaxies in filaments. Instead, the GAEA model includes the assembly bias in dense surroundings and the interaction with host halos for satellite galaxies~\citep{De+Lucia+2024}. 
\par
As 
schematically illustrated in Fig.~\ref{fig:scheme}, filaments can contain both galaxy groups and isolated galaxies~(alone in their halo). In GAEA-all, which includes galaxies with $\log \rm M_{\star} > 8.3$, 51$\pm$1\% of the filament members are simultaneously group members~(17$\pm$1\%, 18$\pm$1\% and 14$\pm$1\% for 1 < N$_{\rm{mem}}$ < 5, 5 < N$_{\rm{mem}}$ < 15, 15 < N$_{\rm{mem}}$ groups, respectively), while 49$\pm$1\%\footnote{This fraction does not depend on the selected persistence level.} of the total filament population are galaxies in isolation -- from now on  ``pure filament galaxies''\footnote{Pure filament galaxies does not belong to any group according to the definition given in Sect.~\ref{subsec:env_idn}}. None of the two populations are negligible. \textcolor{black}{These results were obtained taking into account the selection function, so the fraction of pure filament galaxies was overestimated. Indeed, \cite{Kuchner+2022} reported only 33\% of pure filament galaxies~(pristine in their nomenclature). } 
\par
We are now in the position of better characterising the role of filaments in galaxy evolution: i) considering galaxies in both groups and filaments, we can investigate whether the presence of  groups in the filaments plays a relevant role and, at a fixed mass of the group, whether members of groups outside and inside the filaments have the same \ce{HI} or \ce{H2} content (Sec \ref{sec:mhalo_dependence}); ii) considering only pure filament galaxies we can investigate the influence of filaments on the evolution of galaxies, excluding any group contributions (Sect. \ref{sec:impact_fils}).
Indeed, in GAEA, pure filament galaxies are, by construction, treated similarly to pure field galaxies~(single galaxies inside their haloes). According to our definition of galaxy environment, the only distinction between pure field and pure filament galaxies is their distance from the filaments, with the former being more than 2Mpc/h away from a filament axis and the latter being closer.

\subsection{Dependence on halo mass}
\label{sec:mhalo_dependence}

We investigate whether the filaments have an impact on the galaxies within haloes of fixed mass. To do this, we compare the deficiency of atomic and molecular hydrogen in halos of equal mass inside and outside the filaments.  In order to avoid any biases related to the fact that massive haloes are more prevalent inside filaments while low-mass haloes are more dominant outside filaments~\citep{Welker+2018}, we first fit the $\rm{M}_{halo}$ distribution for haloes inside and outside filaments. 


\begin{figure}
    \centering
    \includegraphics[width=1\linewidth]{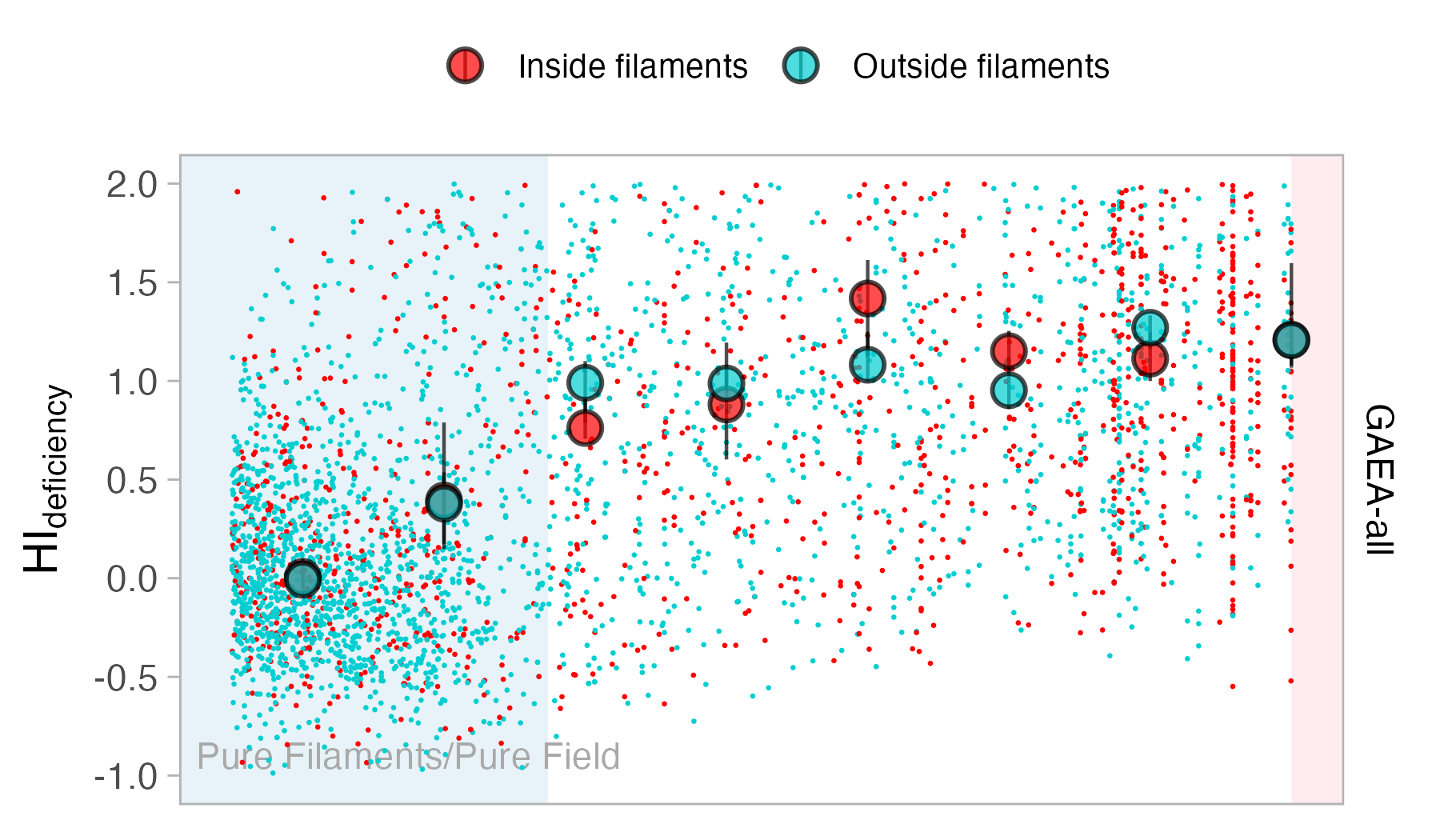}
    \includegraphics[width=1\linewidth]{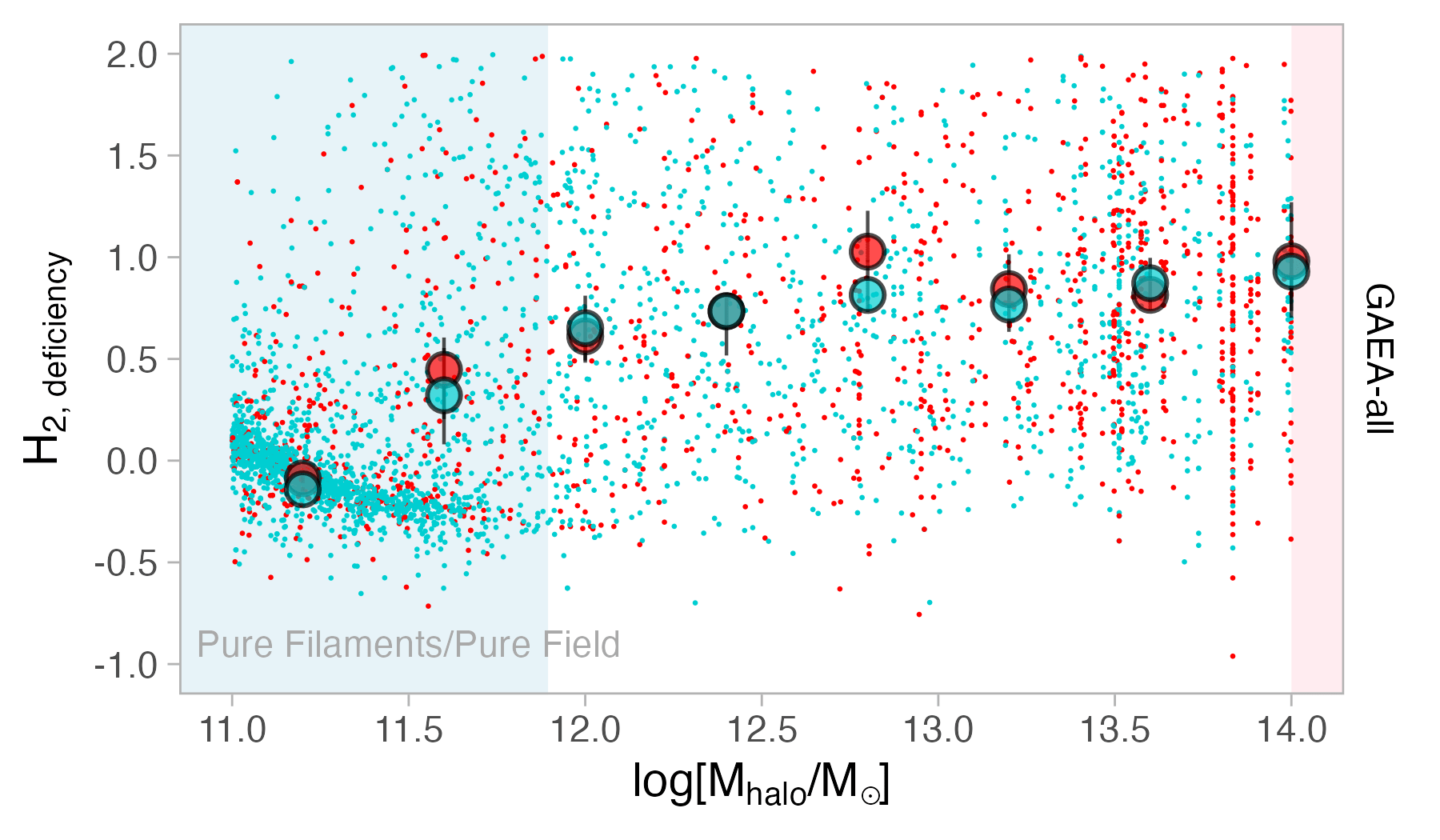}
    \caption{The dependency of atomic and molecular content at fixed halo mass for galaxies inside and outside filaments.
    Top: \ce{HI}-deficiency as a function of the host halo mass $\rm{M}_{halo}$ in GAEA-all. Colour coding reflects the position concerning filaments: inside or outside. Big circles show median values with 1$\sigma$ confidential interval for galaxies inside and outside filaments in each mass bin~(median values were estimated for similar $\rm{M}_{halo}$ distribution for haloes inside and outside filaments by bootstrapping). On the background, typical pure filament/pure field~(90\%-quantile) and clusters of halo masses are highlighted. Bottom: Same but for \ce{H2}-deficiency.}
    \label{fig:mhidef_mhalo}
\end{figure}
\par
Figure~\ref{fig:mhidef_mhalo} shows the \ce{HI}-deficiency and \ce{H2}-deficiency as a function of the host halo mass $\rm{M}_{halo}$ for GAEA-all separately for galaxies inside and outside filaments. The figure presents the median value with 1$\sigma$ significance interval for the galaxies inside and outside filaments within $\rm{M}_{halo}$ bins.
\par
Overall, the median \ce{HI}-deficiency and \ce{H2}-deficiency monotonously increase with the increasing host halo mass.
The median \ce{HI}- or \ce{H2}-deficiency for a given $\rm{M}_{halo}$ is the same within errors for galaxies inside and outside filaments. Thus,  galaxies inside groups with  $\rm{M}_{halo} > 10^{12} \rm{M}_{\odot}$, either inside and outside filaments, have the same \ce{HI} and \ce{H2}-deficiency; that is, the location of the group inside or outside the filaments does not have a significant effect on the amount of cold gas in galaxies. Thus, we do not detect a difference between the deficit of atomic and molecular hydrogen in satellites in groups (the proportion of central galaxies in the considered halos  $\rm{M}_{halo} > 10^{12} \rm{M}_{\odot}$ is 3 percent) inside and outside the filaments, but we note that \cite{Poudel+2017} detects a difference between the central galaxies.
\par
Figure~\ref{fig:mhidef_mhalo} also allowed us to directly compare galaxies inside and outside the filaments (which we classified as pure filament and pure fields, respectively). 
Since up to 90\% of isolated galaxies are located within low-mass haloes $\rm{M}_{halo} < 10^{12} \rm{M}_{\odot}$, these objects populate the light blue/grey shaded area in Fig.~\ref{fig:mhidef_mhalo}. For both \ce{HI}- and \ce{H2}-deficiency, at fixed halo mass isolated galaxies inside and outside filaments have a comparable deficiency of atomic or molecular hydrogen. Therefore, we do not detect any signs of the role of filaments  on \ce{HI}- or \ce{H2}-content for pure filament members.


\subsection{The impact of filaments  on the galaxy \ce{HI} and \ce{H2} content}
\label{sec:impact_fils}


\begin{figure*}
    \centering
    {\includegraphics[width=1\textwidth]{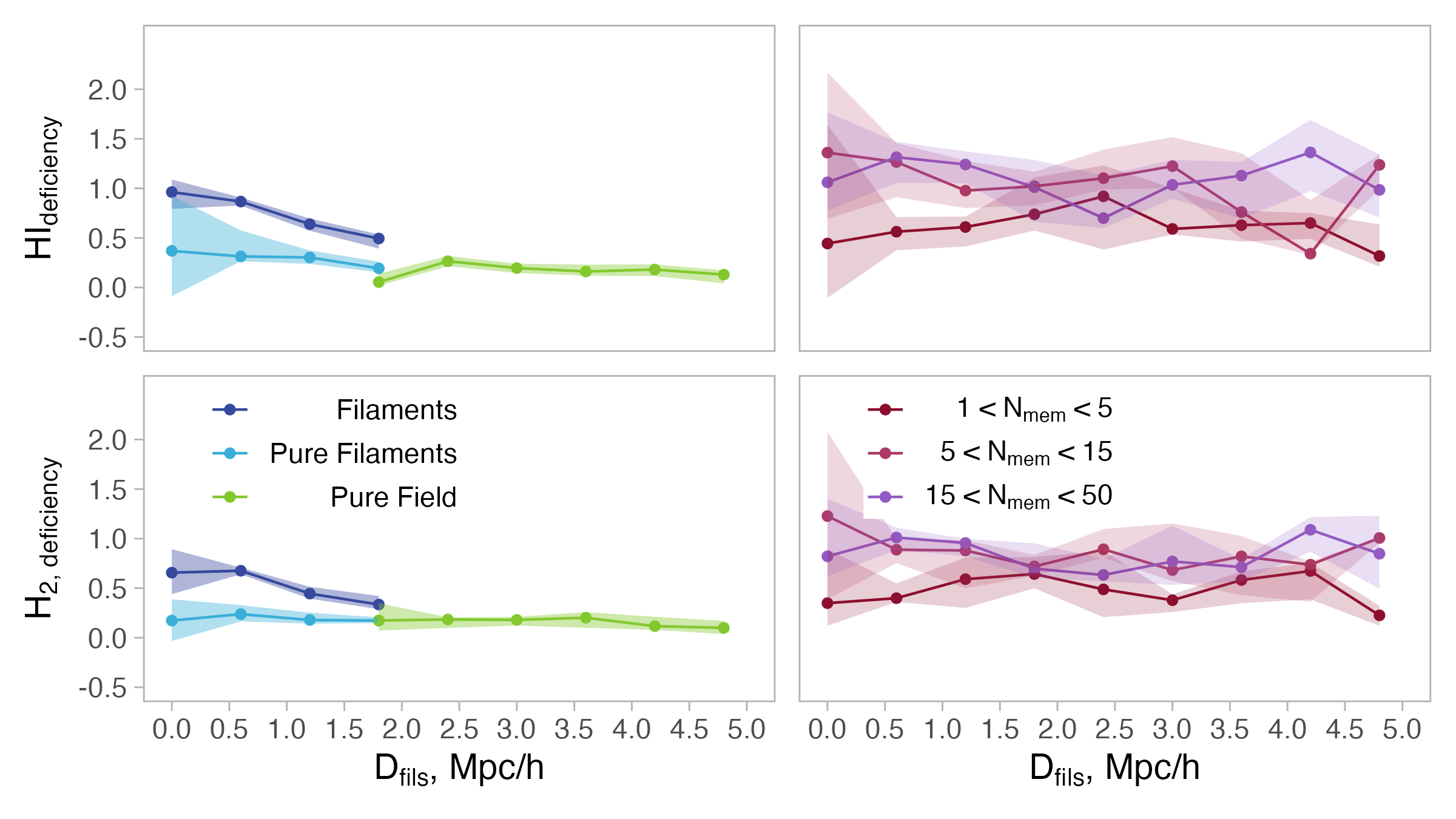}}
    {\caption{\ce{HI}- or \ce{H2}-deficiency (median values and 1$\sigma$ significance interval) as a function of 3D distance to the nearest filament in GAEA-all for filaments, the pure filament, and pure fields~ (top panels), and members of groups of different sizes: 1 < N$_{\rm{mem}}$ < 5, 
     5 < N$_{\rm{mem}}$ < 15, 15 < N$_{\rm{mem}}$ < 50~(bottom panels).}
    \label{fig:mhidef_mh2def_disttoggfs_distored}}
\end{figure*}


The most common method for determining the influence of filaments on galaxies is to check the dependence of their properties (mass/star formation rate/amount of gas) as a function of the distance to the filaments~\citep[e.g.,][]{Singh+2020,Hasan+2023}. Observations have shown that the influence of filaments on galaxy properties is usually more pronounced near the filaments axis: galaxy in filaments are typically redder~\citep{Singh+2020}, more HI-deficient~\citep{Odekon+2018, Lee+2021}, more massive~\citep{Kraljic+2018} and have earlier morphological types~\citep{Castignani+2022_gas} if they lie closer the filament axis. 

Following the same approach, for each galaxy from GAEA-all, we determine the 3D distance to its nearest filament. Next, we define the \ce{HI}- and \ce{H2}-deficiency as a function of distance to filament for galaxies in groups, filaments, pure filament, and pure field separately.
Results are presented in Fig.~\ref{fig:mhidef_mh2def_disttoggfs_distored}.
In agreement with previous works~\citep{Castignani+2022_gas, Odekon+2018,Lee+2021,Hoosain+2024},
filament members close to the filament axis are more \ce{HI}- and \ce{H2}-deficient that in the outer parts of the filaments (top panels): \ce{HI}$_{\rm{def}}$=0.96$\pm$0.17 and \ce{H2}$_{, \rm{def}}$=0.66$\pm$0.2 at $\rm{D}_{\rm{fil}} < 0.5$Mpc/h vs \ce{HI}$_{\rm{def}}$=0.48$\pm$0.07 and \ce{H2}$_{, \rm{def}}$=0.33$\pm$0.06 at $ 1.5 < \rm{D}_{\rm{fil}} < 2.0$Mpc/h. In contrast, pure filament galaxies show much lower absolute values of both \ce{HI}- and \ce{H2}-deficiency~(\ce{HI}$_{\rm{def}}$=0.35$\pm$0.5 and \ce{H2}$_{, \rm{def}}$=0.17$\pm$0.2 at $\rm{D}_{\rm{fil}} < 0.5$Mpc/h) and almost no dependence on the distance to filament. Moreover, 
pure field galaxies shows the same properties of \ce{HI}- and \ce{H2}-deficiency as  pure filament galaxies. This emphasizes that these galaxies essentially have similar properties of atomic and molecular hydrogen. 
\par
The right panels of Fig.~\ref{fig:mhidef_mh2def_disttoggfs_distored} consider  members of groups of different sizes: 1 < $\rm N_{\rm mem}$ < 5, 5 < $\rm N_{\rm mem}$ < 15, and 15 < $\rm N_{\rm mem}$ < 50\footnote{The number of group members corresponds to the number of galaxies with $\log_{10} [{\rm M}_{\star} / \rm{M}_{\sun} ] > 8.3$ in the halo after mimicking the selection function~(see Sect.~\ref{subsec:masscut_83}).}. Galaxies in groups closer than 2 Mpc/h from the filaments axis are also members of the filaments. Overall,  group galaxies inside and outdside filaments are more  \ce{HI}- and \ce{H2}-deficient than pure filament galaxies, regardless of the group richness. 1 < $\rm N_{\rm mem}$ < 5 groups statistically show lower \ce{HI}- and \ce{H2}-deficiencies than bigger groups~(5 < $\rm N_{\rm mem}$ < 15, and 15 < $\rm N_{\rm mem}$ < 50) and do not show a dependence on the proximity to filaments. In contrast,  groups with 5 < $\rm N_{\rm mem}$ < 15 and 15 < $\rm N_{\rm mem}$ < 50 show a higher \ce{HI}- and \ce{H2}- deficiency close to the filament axis: \ce{HI}$_{\rm{def}}$=1.36$\pm$0.74 and \ce{H2}$_{, \rm{def}}$=1.23$\pm$0.85 at $\rm{D}_{\rm{fil}} < 0.5$Mpc/h vs \ce{HI}$_{\rm{def}}$=1.07$\pm$0.16 and \ce{H2}$_{, \rm{def}}$=0.72$\pm$0.11 at $ 1.5 < \rm{D}_{\rm{fil}} < 2.0$Mpc/h in groups  5 < N$_{\rm{mem}}$ < 15~(\ce{HI}$_{\rm{def}}$=1.06$\pm$0.4 and \ce{H2}$_{, \rm{def}}$=0.85$\pm$0.35 at $\rm{D}_{\rm{fil}} < 0.5$Mpc/h vs \ce{HI}$_{\rm{def}}$=1.01$\pm$0.3 and \ce{H2}$_{, \rm{def}}$=0.7$\pm$0.12 at $ 1.5 < \rm{D}_{\rm{fil}} < 2.0$Mpc/h in groups  15 < N$_{\rm{mem}}$ < 50~). \textcolor{black}{We note that all the dependencies given in this section are calculated relative to filaments, taking into account peculiar velocity distortions~(see Appendix~\ref{appendix:impact_of_elongation} for the details). We discuss the impact of the line-of-sights distortions on the filament extraction on this test in Appendix~\ref{app:test_check}.}

\par
As a consequence, the result of filament galaxies having  gas content intermediate 
between the cluster and the pure field, shown in Sect.~\ref{sec:results} and in \cite{Castignani+2022_gas}, can be explained by the fact that filaments host both galaxies in groups, and pure filament galaxies, in similar proportions and that the two populations have different  \ce{HI} and \ce{H2} properties.

Besides, \cite{Hoosain+2024} 
found in the RESOLVE survey~\citep{Stark+2016} and the ECO catalogue~\citep{Eckert+2017} that compared group galaxies and isolated systems inside filaments, they also reveal that tendency of overall filaments members to be more \ce{HI} and \ce{H2}-deficient~(i.e. intermediate properties) near the filament axis relate to the increasing roles of galaxy groups inside filaments rather than isolated galaxies~(isolated galaxies close do not demonstrate dependency of \ce{HI} and \ce{H2}-deficiency on the distance to filaments).

\par

Our statement that galaxies in pure filament have similar properties to galaxies in the pure field is based mainly on the fact that the dependence of the \ce{HI}- and \ce{H2}-deficiency does not depend on the distance to the filaments. However, our filament structure was calculated taking into account elongation along line-of-sight effects, which did not allow us to accurately estimate the distance to the filament axis~(see Appendix~\ref{appendix:impact_of_elongation}) and can consequently distort the results of this test.
Therefore, we repeat the same exercise, after having redefined the environment and distance to filaments for each galaxy from GAEA-all relative to the `true filaments`. We call `true filaments` those determined by DiSPerSE using the distribution of all galaxies with $\log_{10} [{\rm M}_{\star} / \rm{M}_{\sun} ] > 8.3$ in the cartesian coordinates x-y-z of the model around Virgo-like clusters with a persistence level of 4$\sigma$~(as described in  Appendix~\ref{appendix:impact_of_elongation}). We also consider as pure filament galaxies only those who are truly isolated in the model; that is, there are no other gravitationally bounded galaxies of any mass, and no selection function applied.  Also in this case,  we do not find any dependency of \ce{HI}- or \ce{H2}-deficiency on the distance to filaments, reassuring us about the robustness of our results. 
\par
Nonetheless, it is important to keep in mind that the adopted selection function might impact the results: about 33$\pm$2\% of galaxies that appear as pure filament galaxies in our sample are actually members of groups. In these cases, the observed dependence of \ce{HI}- or \ce{H2}-deficiency on the distance to filaments (as was also shown in Fig.7 of \cite{Odekon+2018}), might simply be due to the fact that we are actually considering group members, for which the trend is clearly established. 
\par
Thus, the GAEA model does not predict any filament influence on the \ce{HI}- or \ce{H2}-deficiency at fixed halo mass.

\section{Conclusions}
\label{sec:conclusion}

The main goal of this paper was to investigate whether the GAEA semi-analytic model, which has explicit prescriptions for partitioning the cold gas content in its atomic and molecular phases, is able to reproduce the observational results of \citep{Castignani+2022_gas}, who characterised the gas content of galaxies in the filaments surrounding the Virgo cluster. 
To that end, we carefully extracted from the model, samples of galaxies to best mimic the observational data, including some selection biases. We extracted filaments surrounding clusters with a mass similar to that of Virgo; we applied the observational mass and \ce{HI} and \ce{H2} completeness limits; and we defined environments in a homogeneous way in the observations and the model. 
In addition, we extracted from the model a sample of galaxies not affected by observational biases in order to make more general statements about the predictions of the models on the gas content of galaxies. The main findings of this work are summarised as follows:

\begin{enumerate}
    \item When considering \ce{HI}, the model is able to reproduce very well the observational results. The observed and model data have similar $\rm{M}_{\ce{HI}}$-$\rm{M}_{\star}$ relations. The fraction of \ce{HI}-deficient galaxies decreases from clusters to filaments and to the pure field and increases with increasing stellar mass, as was shown for regions beyond the Virgo cluster~\citep{Denes+2016, Odekon+2018,Zabel+2019}. The only exception is in Virgo, where it is consistent with being flat across the considered mass range. In the regime where no observations are available, the model predicts a larger  \ce{HI}-deficiency for low-mass galaxies than for massive ones. GAEA is able to reproduce not only the \ce{HI}-deficient fraction but also the observed cumulative distribution function of the \ce{HI}-deficiency at all masses.

    \item Focusing on \ce{H2}, we also observed a $\rm{M}_{\ce{H2}}$-$\rm{M}_{\star}$  correlation for all the considered environments, a good agreement between the observations and the model, and an enhancement of \ce{H2}-deficient galaxies in cluster and filament galaxies with respect to the field and among massive galaxies. In the regime where no observations are available, the model predicts a larger  \ce{H2}-deficiency in clusters, in agreement with the Coma~\citep{Casoli+1991} and Fornax~\citep{Zabel+2019} clusters. GAEA also reproduces the observed cumulative distribution function of the \ce{H2}-deficiency at all masses. 

    \item In both the observations and the model, we find a correlation between \ce{H2}-deficiency and \ce{HI}-deficiency for each environment and mass bin, as was shown in \cite{Zabel+2022}. Low-mass galaxies are mostly both \ce{HI} and \ce{H2} normal and star-forming. GAEA, however, predicts a larger fraction of \ce{HI}- and \ce{H2}-deficient galaxies~(63$\pm$2\%) than in other environments~(48$\pm$1\% and 18$\pm$1\% of filaments and pure field galaxies). In contrast, high-mass galaxies are either both \ce{HI}- and \ce{H2}-normal or both \ce{HI} and \ce{H2}-deficient. The fraction of galaxies deficient in only one of the gas phases is lower than 20\%, according to the model. Overall, the amount of atomic hydrogen \ce{HI} is more sensitive to the environment than molecular hydrogen \ce{H2}, in agreement with many findings.

    \item Taking into account all possible observational biases, the GAEA-mock reproduces the observed \ce{HI} and \ce{H2} deficiencies in galaxies in clusters, filaments, and fields, even if intrinsic relations from GAEA-all are different (for the low-mass end).

\end{enumerate}

Our analysis therefore confirms the results by \cite{Castignani+2022_gas} and \cite{Castignani+2022_catalogue} that filaments have intermediate properties between cluster and field galaxies also from a theoretical point of view. We stress that the model does not include any special processing of the filaments. 

We can explain the intermediate properties of filaments by taking into account the fact that they consist of ~50\% isolated galaxies, which have properties similar to pure field galaxies, and~50\% group members, which have gas properties similar to those of a cluster. In the model, the \ce{HI}- and \ce{H2}-deficiency of isolated galaxies in filaments does not depend on the actual distance to the filaments, which means similar assembly histories for isolated galaxies inside and outside the filaments. Similarly, the fact that galaxies in groups inside or outside filaments have similar properties suggests that filaments themselves are not able to strongly impact the gas content. However, this does not exclude the role of filaments in the gaseous evolution of galaxies, but we expect it to be a second order effect. In addition, in this paper, we focused on the gas component, which is not an integral parameter over time. Other properties such as stellar mass and SFR may more significantly depend on the environment. We aim to delve deeper into these aspects  in an upcoming paper in this series.

We have shown that low-mass galaxies ($\rm{M}_{\star} < 10^{10} \rm{M}_{\odot}$) are more sensitive to environmental effects than massive ones ($\rm{M}_{\star} > 10^{10} \rm{M}_{\odot}$). \cite{Donnari+2021} have shown the same for the hydrodynamical simulation IllustrisTNG~\citep{Illustris1,Illustris2,Illustris3,Illustris4,Illustris5} by demonstrating that low-mass galaxies experience environmental quenching when massive galaxies quench on their own~(AGN-feedback).  Future surveys such as WEAVE~\citep{WEAVE}, WALLABY~\citep{SKA_WALLABY_2020}, and MIGHTEE-HI~\citep{MIGHTEE+2021}, which will provide large statistical samples of  low-mass galaxies, will be particularly important in confirming our predictions and establishing the role of the environment in galaxy evolution.


\begin{acknowledgements}
    \textcolor{black}{We thank the anonymous referee for the comments, which helped improve the quality of the presentation of our results. }  
    \par
      DZ and BV acknowledge support from the INAF Mini Grant 2022 “Tracing filaments through cosmic time” (PI Vulcani).  The authors thank Olga Cucciati for useful discussions on the coordinate transformations. GC acknowledges the support from the Next Generation EU funds within the National Recovery and Resilience Plan (PNRR), Mission 4 - Education and Research, Component 2 - From Research to Business (M4C2), Investment Line 3.1 - Strengthening and creation of Research Infrastructures, Project IR0000012 – “CTA+ - Cherenkov Telescope Array Plus”. GHR acknowledges the support of NASA ADAP grant 80NSSC21K0641, and NSF AAG grants AST-1716690 and AST-2308127. GHR also acknowledges the hospitality of Hamburg Observatory, who hosted him during parts of this work. 
        \par
      The Legacy Surveys consist of three individual and complementary projects: the Dark Energy Camera Legacy Survey (DECaLS; Proposal ID \#2014B-0404; PIs: David Schlegel and Arjun Dey), the Beijing-Arizona Sky Survey (BASS; NOAO Prop. ID \#2015A-0801; PIs: Zhou Xu and Xiaohui Fan), and the Mayall z-band Legacy Survey (MzLS; Prop. ID \#2016A-0453; PI: Arjun Dey). DECaLS, BASS and MzLS together include data obtained, respectively, at the Blanco telescope, Cerro Tololo Inter-American Observatory, NSF’s NOIRLab; the Bok telescope, Steward Observatory, University of Arizona; and the Mayall telescope, Kitt Peak National Observatory, NOIRLab. Pipeline processing and analyses of the data were supported by NOIRLab and the Lawrence Berkeley National Laboratory (LBNL). The Legacy Surveys project is honored to be permitted to conduct astronomical research on Iolkam Du’ag (Kitt Peak), a mountain with particular significance to the Tohono O’odham Nation.

    NOIRLab is operated by the Association of Universities for Research in Astronomy (AURA) under a cooperative agreement with the National Science Foundation. LBNL is managed by the Regents of the University of California under contract to the U.S. Department of Energy.
    
    This project used data obtained with the Dark Energy Camera (DECam), which was constructed by the Dark Energy Survey (DES) collaboration. Funding for the DES Projects has been provided by the U.S. Department of Energy, the U.S. National Science Foundation, the Ministry of Science and Education of Spain, the Science and Technology Facilities Council of the United Kingdom, the Higher Education Funding Council for England, the National Center for Supercomputing Applications at the University of Illinois at Urbana-Champaign, the Kavli Institute of Cosmological Physics at the University of Chicago, Center for Cosmology and Astro-Particle Physics at the Ohio State University, the Mitchell Institute for Fundamental Physics and Astronomy at Texas A\&M University, Financiadora de Estudos e Projetos, Fundacao Carlos Chagas Filho de Amparo, Financiadora de Estudos e Projetos, Fundacao Carlos Chagas Filho de Amparo a Pesquisa do Estado do Rio de Janeiro, Conselho Nacional de Desenvolvimento Cientifico e Tecnologico and the Ministerio da Ciencia, Tecnologia e Inovacao, the Deutsche Forschungsgemeinschaft and the Collaborating Institutions in the Dark Energy Survey. The Collaborating Institutions are Argonne National Laboratory, the University of California at Santa Cruz, the University of Cambridge, Centro de Investigaciones Energeticas, Medioambientales y Tecnologicas-Madrid, the University of Chicago, University College London, the DES-Brazil Consortium, the University of Edinburgh, the Eidgenossische Technische Hochschule (ETH) Zurich, Fermi National Accelerator Laboratory, the University of Illinois at Urbana-Champaign, the Institut de Ciencies de l’Espai (IEEC/CSIC), the Institut de Fisica d’Altes Energies, Lawrence Berkeley National Laboratory, the Ludwig Maximilians Universitat Munchen and the associated Excellence Cluster Universe, the University of Michigan, NSF’s NOIRLab, the University of Nottingham, the Ohio State University, the University of Pennsylvania, the University of Portsmouth, SLAC National Accelerator Laboratory, Stanford University, the University of Sussex, and Texas A\&M University.
    
    BASS is a key project of the Telescope Access Program (TAP), which has been funded by the National Astronomical Observatories of China, the Chinese Academy of Sciences (the Strategic Priority Research Program “The Emergence of Cosmological Structures” Grant \# XDB09000000), and the Special Fund for Astronomy from the Ministry of Finance. The BASS is also supported by the External Cooperation Program of Chinese Academy of Sciences (Grant \# 114A11KYSB20160057), and Chinese National Natural Science Foundation (Grant \# 12120101003, \# 11433005).
    
    The Legacy Survey team makes use of data products from the Near-Earth Object Wide-field Infrared Survey Explorer (NEOWISE), which is a project of the Jet Propulsion Laboratory/California Institute of Technology. NEOWISE is funded by the National Aeronautics and Space Administration.
    
    The Legacy Surveys imaging of the DESI footprint is supported by the Director, Office of Science, Office of High Energy Physics of the U.S. Department of Energy under Contract No. DE-AC02-05CH1123, by the National Energy Research Scientific Computing Center, a DOE Office of Science User Facility under the same contract; and by the U.S. National Science Foundation, Division of Astronomical Sciences under Contract No. AST-0950945 to NOAO.

    The Siena Galaxy Atlas was made possible by funding support from the U.S. Department of Energy, Office of Science, Office of High Energy Physics under Award Number DE-SC0020086 and from the National Science Foundation under grant AST-1616414.
\end{acknowledgements}

%
  \bibliographystyle{aa520} 
  \bibliography{main_aa} 
%
\begin{appendix}


\section{Coordinate transformation from x-y-z to RA-DEC-z to SGX-SGY-SGZ}
\label{app:coord_transform}

To mimic the distortion of the cosmic web due to the elongation along the line-of-sight as in the Virgo cluster~\cite{Tully+2008}, we transformed the Cartesian x-y-z coordinates in the GAEA model to RA-DEC-z coordinates. 
\par
We first put the position of the pseudo-observer at the same distance from the target halo of that of the Virgo cluster. Next, we computed the position of each galaxy in the entire GAEA cube relative to the pseudo-observer:
\begin{equation}
    \begin{cases}
      x^{'} = x - x_{observer},  \\
      y^{'} = y - y_{observer}, \\
      z^{'} = z - z_{observer} + D_{snapshot},
      
    \end{cases}\,,
\end{equation}
where $D_{snapshot}$ is the distance to the snapshot (only for cases  where redshift $\mathrm{z}$ > 0). For each galaxy, we estimated the co-moving distance to the pseudo-observer at  $\mathrm{z}$ = 0:
\begin{equation}
    D_{comov} = \sqrt{( x^{' 2} + y^{' 2} +z^{' 2})},
\end{equation}
and to transform it to over h: $D_{comov} = D_{comov}\cdot 100/H_{0}$. Using these distances, we calculated the redshift $\mathrm{z}_{cos}$ for each galaxy relative to the pseudo-observer by interpolation. 

The peculiar velocities were calculated according to the formula\begin{equation}
    v_{pec} = ( x^{'} v_{x} + y^{'} v_{y} +
               z^{'} v_{z}) / D_{comov}.
\end{equation}Next, we corrected the redshift, taking the peculiar velocities into account:
\begin{equation}
    \mathrm{z}_{obs} = \mathrm{z}_{cos} + v_{pec}/c + \mathrm{z}_{cos} v_{pec}/c,
\end{equation}
then 
\begin{equation}
    \begin{cases}
        RA = atan(y^{'} / z^{'})\\
       DEC = asin(x^{'} / D_{comov})
      
    \end{cases}\,
.\end{equation}
To identify an area of similar volume to \cite{Castignani+2022_catalogue}, we also evaluated the recession velocities:

\begin{equation}
    v_{r} = D_{comov} \cdot H_{0} + v_{pec}
.\end{equation}
Using (RA, DEC, $\mathrm{z}_{obs}$), we computed supergalactic spherical coordinates (SGL, SGB).  To convert (SGL, SGB) to (SGX, SGY, SGZ):
\begin{equation}
    \begin{cases}
        SGX = D_{vr} \sin(SGL) \cos(SGB) \\
        SGY = D_{vr} \sin(SGL) \sin(SGB) \\
        SGZ = D_{vr} \cos(SGL), \\
      
    \end{cases}\,
\end{equation}
where $D_{vr} = v_{r} / H_0$. 


\section{The impact of the line-of-sight elongation on the identification of filaments}
\label{appendix:impact_of_elongation}


\begin{figure}
    \centering
    \includegraphics[width=1\linewidth]{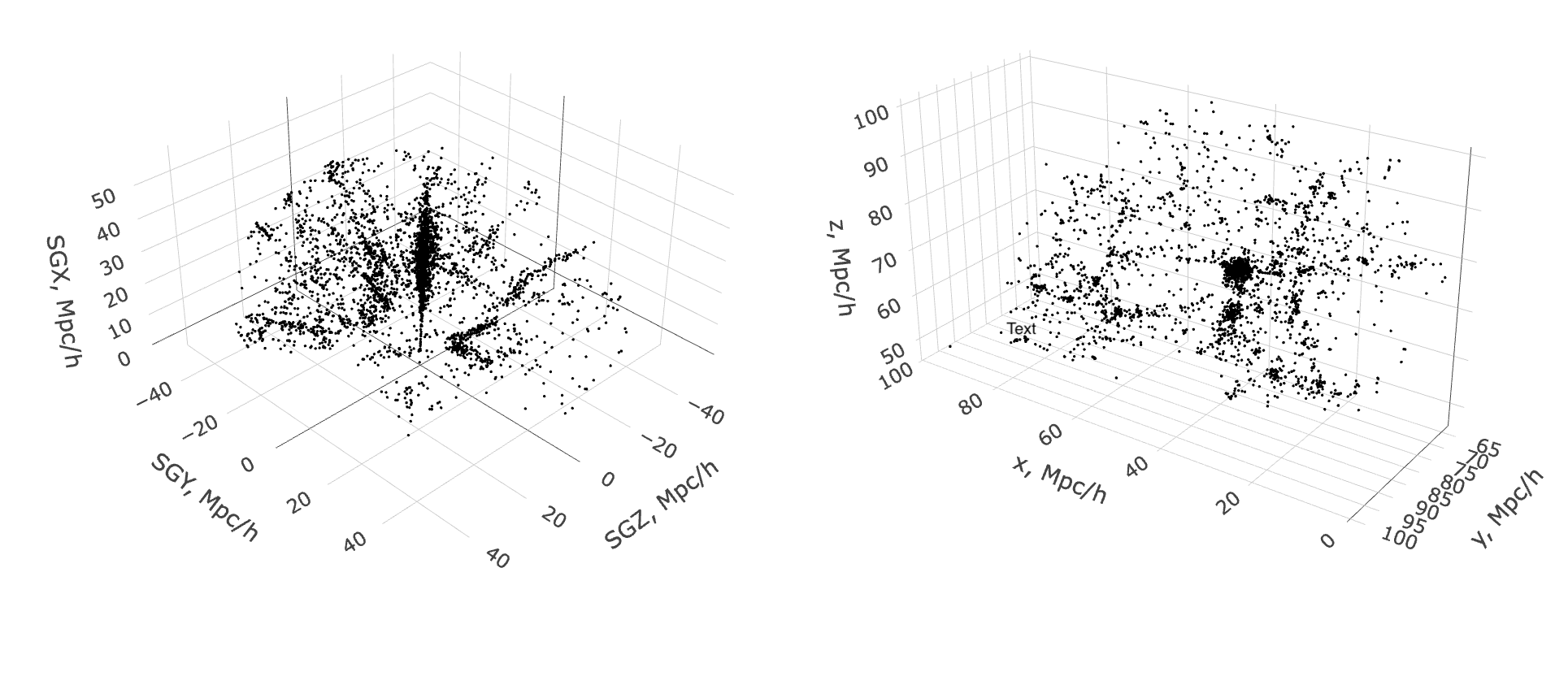}
    \caption{Illustration of how supergalactic coordinates distort the true position of galaxies. Here we show the distribution of exactly the same galaxies as in GAEA V2 but in a different coordinates system; the same area in the RA-DEC plane is shown in Fig.~\ref{fig:radec_filsmems}~(bottom-left panel). Left:  Distribution of galaxies in supergalactic SGX-SGY-SGZ coordinates~(includes elongation along the line of sight). Right: Same as left but in GAEA coordinate system x-y-z. }
    \label{fig:example_of_coords}
\end{figure}

Previous works in the series \citep{Castignani+2022_gas, Castignani+2022_catalogue} examined 3D filaments in supergalactic coordinates~\citep{Tully+1982}. For consistency with those papers, in this study we also investigated filaments in supergalactic coordinates, which includes peculiar velocity effects. This changes the topology of the distribution of galaxies and impacts the filament identification. For instance, the finger-of-god~(FoG) elongation leads to unreliable filament determinations~\citep{Kraljic+2018, Kuchner+2021}. Figure~\ref{fig:example_of_coords} demonstrates the disparity between galaxies' distribution in supergalactic SGX-SGY-SGZ and coordinates from the model.

In this Appendix, we check how much the filaments extracted from a galaxy distribution close to the observed one differ from the filaments obtained from the original galaxy distribution (without line-of-sight effects, in the Cartesian coordinates x-y-z of the model and without mimicking the selection function). We call the last filament structure (FS) `true'. We call filaments extracted by the distribution of galaxies close to observational ones `distorted'.
\par
For GAEA V1, GAEA V2, and GAEA V3 we separately identified not only the filaments described in Sect.~\ref{subsec:filaments_idn} but also a `true filaments system' via the distribution of all galaxies with $\log_{10} [{\rm M}_{\star} / \rm{M}_{\sun} ] > 8.3$~(e.g. without consideration of the selection function, the number of samples is significantly higher) in Cartesian coordinates of model x-y-z with a 4$\sigma$ persistence level. 
\par
First of all, we inspected how much the filament membership determination depends on the way filaments are extracted (true vs distorted). 
This test is critical to obtain and estimate on the error of determining whether the galaxy belongs to the filament. The top panel of Fig.~\ref{fig:car_vs_super} represents the confusion matrix for the classification of galaxies in GAEA-all samples. 18\% of galaxies are members of the filaments and 51\% are non-filament members according to both FSs. A distorted filament structure gives the correct galaxy status in more than 70 percent of cases. The remaining cells show errors in the definition, and their inequality is due to the fact that the true filamentous structure was extracted from a large number of samples with the same level of persistence, which makes it more detailed.
\par
In addition to the status itself~(inside or outside), we are also interested in how much the distorted distribution of galaxies helps restore the exact position of the filament axis. To check this, we calculate for each galaxy from GAEA-all the distance to the nearest filaments in true and distorted FS and compare these distances in the bottom panel of Fig.~\ref{fig:car_vs_super}. Since we do not observe a concentration of points around the line of equality, we conclude that the distorted filaments poorly reflect the true positions of the filament axes and have a significant effect on tests based on the distance to the filaments.

\begin{figure}
    \centering
        \includegraphics[width=1\linewidth]{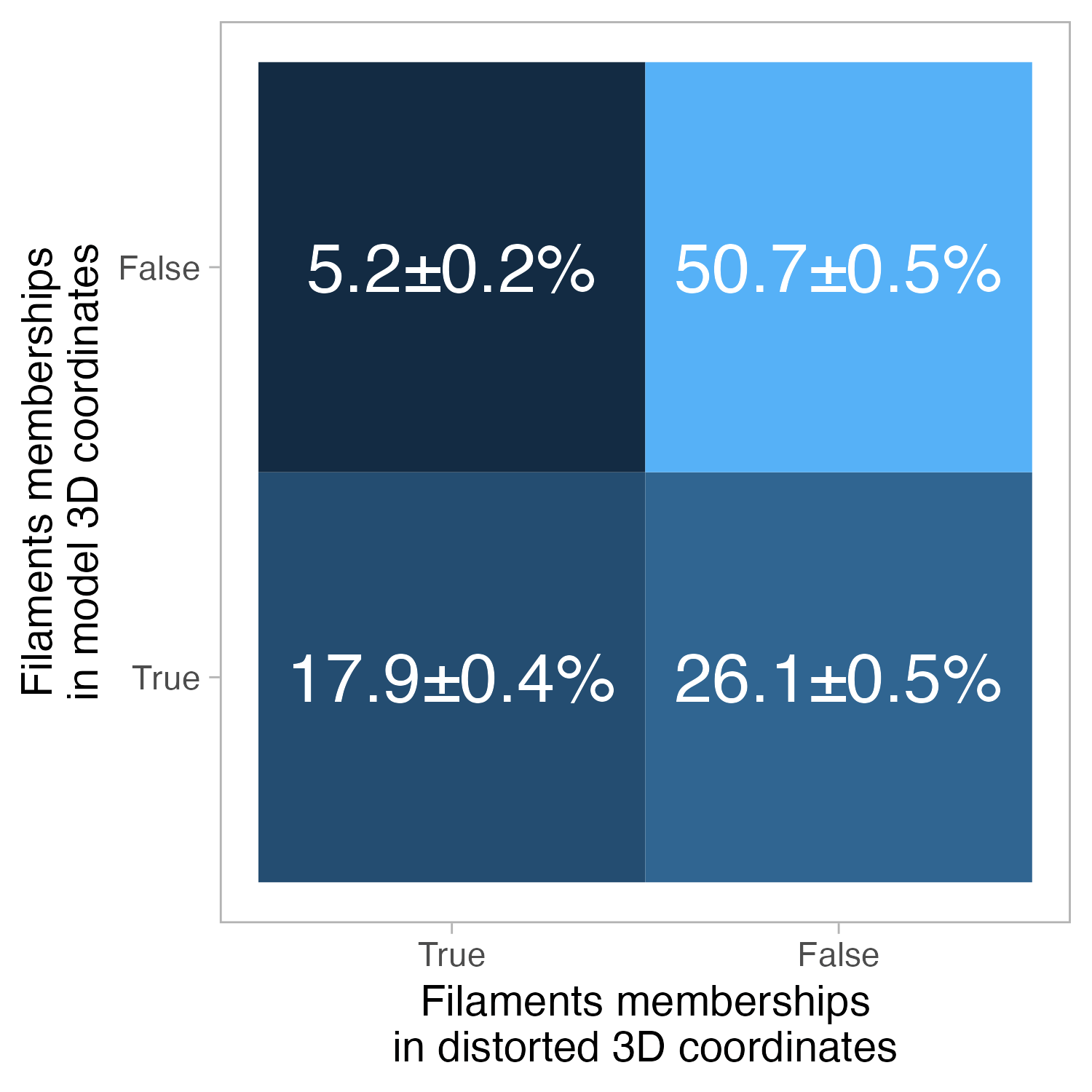}
    \includegraphics[width=1\linewidth]{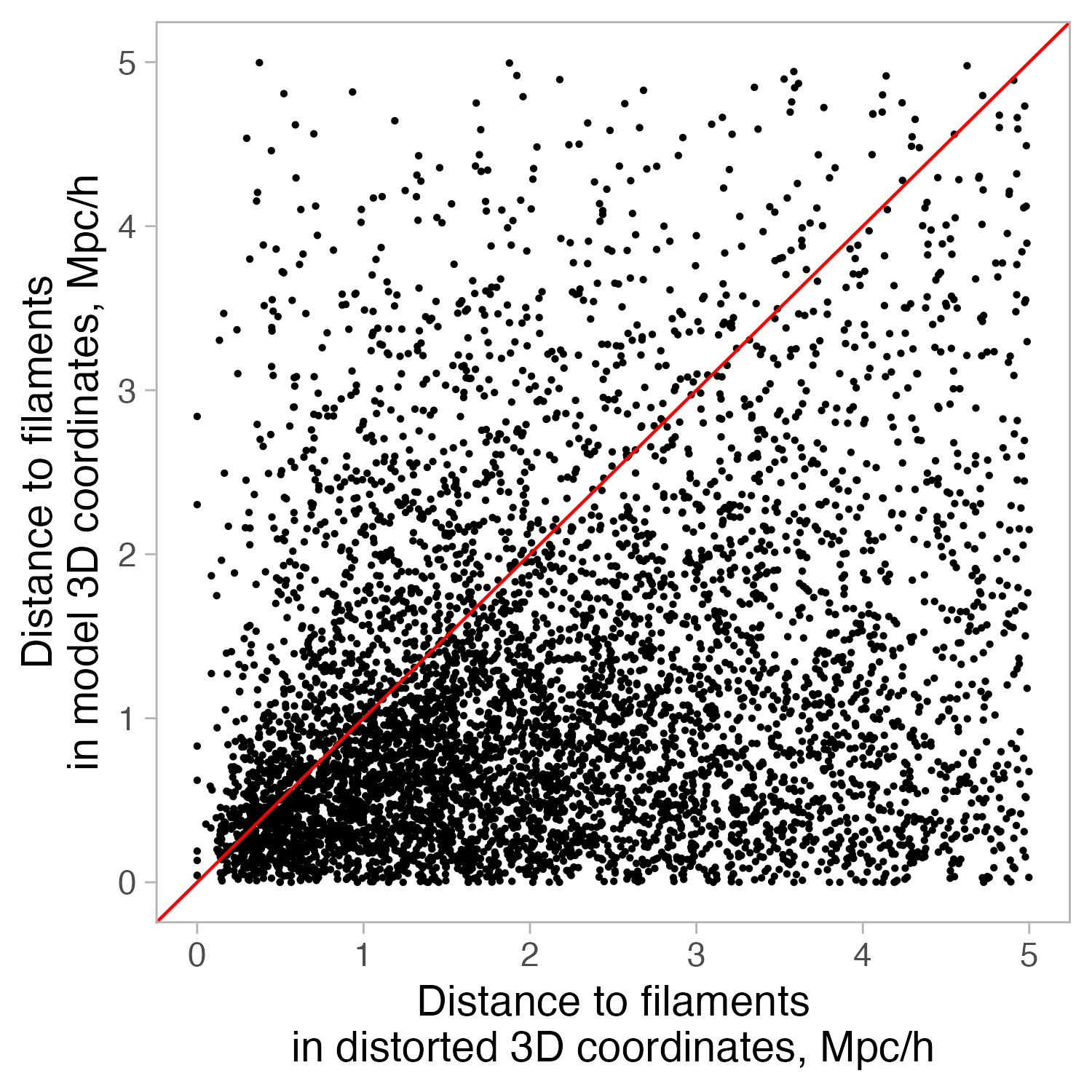}
    \caption{The correspondence between `true` and `distorted` filament identification. Left: Confusion matrix for classification of GAEA-all galaxies as filaments members according to filaments identified in SGX-SGY-SGZ~(see Sect.~\ref{subsec:filaments_idn}) with distortion along line-of-sight (x-axis) and according to filaments identified for GAEA-all without mimicking of the selection function~(see Sect.~\ref{subsec:gaea_coord_transform}) in x-y-z~(`true` filaments, y-axis). The bottom left panel represents the fraction of galaxies identified as filament members according to both filament systems. The right top panel represents the fraction of galaxies identified as field members according to both filaments systems. The left top and bottom right panels represent inconsistency in the classification of filament members.
    Right: Comparison of distances for the indicated filament systems for each galaxy from GAEA-all in 3D. The red line represents the equality line. }
    \label{fig:car_vs_super}
\end{figure}

\par
We additionally note that these conclusions are true for the close-to-observer regions, like Virgo and may differ for more distant ones.


\section{Comparing the \ce{HI}/\ce{H2} deficiency definition}
\label{app:defs_comparison}

For consistency between simulations and observations, in Sect.~\ref{sec:deficiency_defs}, we use a simplified definition of gas content that only includes stellar mass dependence. Here, we compare the obtained values with the values used in \cite{Castignani+2022_catalogue}. Figure~\ref{fig:defs_comparison} shows that the data are in good agreement for \ce{HI} and systematically underestimated values for \ce{H2}. Given the overall small discrepancies, our simplified adopted definition does not negatively affect the analysis.

\begin{figure}
    \centering
    \includegraphics[width=1\linewidth]{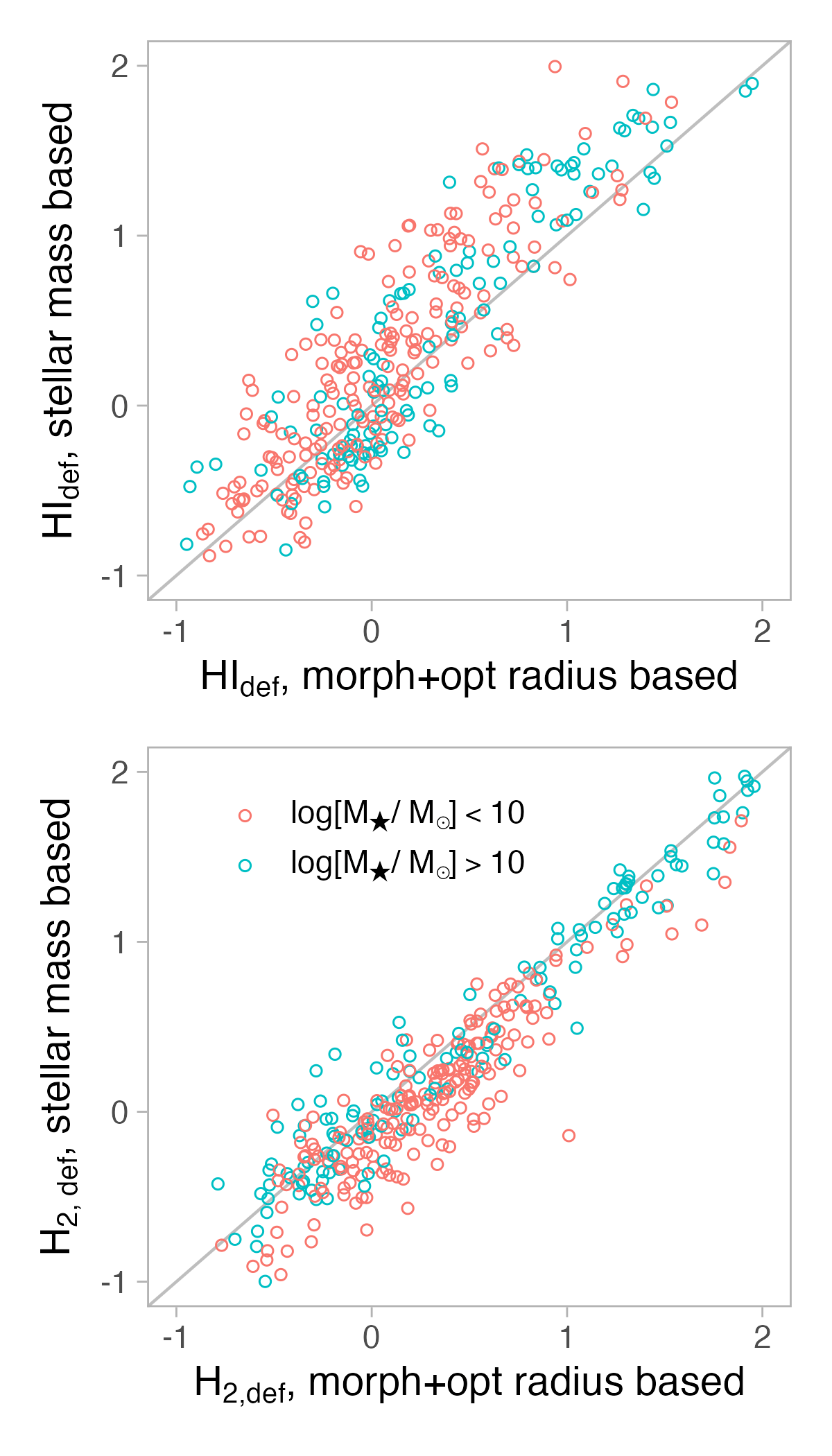}
    \caption{\ce{HI}~(top) and \ce{H2}~(bottom) deficiency defined in this work and in \cite{Castignani+2022_catalogue}. }
    \label{fig:defs_comparison}
\end{figure}

\section{Influence of the line-of-sight effects on the results}
\label{app:test_check}

\textcolor{black}{The effects described in Appendix~\ref{appendix:impact_of_elongation} may also influence the results of this work. In the main paper, we use the distorted filament system~(FS) for a straightforward comparison with observational data. Here we check that our results are not affected when considering the position of filaments without line-of-sight distortions. }
\par
\textcolor{black}{Inf Appendix~\ref{appendix:impact_of_elongation} we have shown that the distorted FS does a good job in classifying galaxies~(inside or outside of filaments) but cannot reflect the exact position of the filament axis. Therefore, we expect that the results of  Fig.~\ref{fig:mhidef_mhalo} should be less affected than the results in  Fig.~\ref{fig:mhidef_mh2def_disttoggfs_distored}.
}
\par
\textcolor{black}{Figure~\ref{app:fig:mhalotest_distored} shows the analogue of the Fig.~\ref{fig:mhidef_mhalo} but considering the 'true' filaments. We again do not recover the difference in \ce{HI} or \ce{H2}-deficiency for galaxies inside or outside filaments with controlled halo mass, so elongation along line-of-sight  does not affect this conclusion. }

\begin{figure}
    \centering
    \includegraphics[width=1\linewidth]{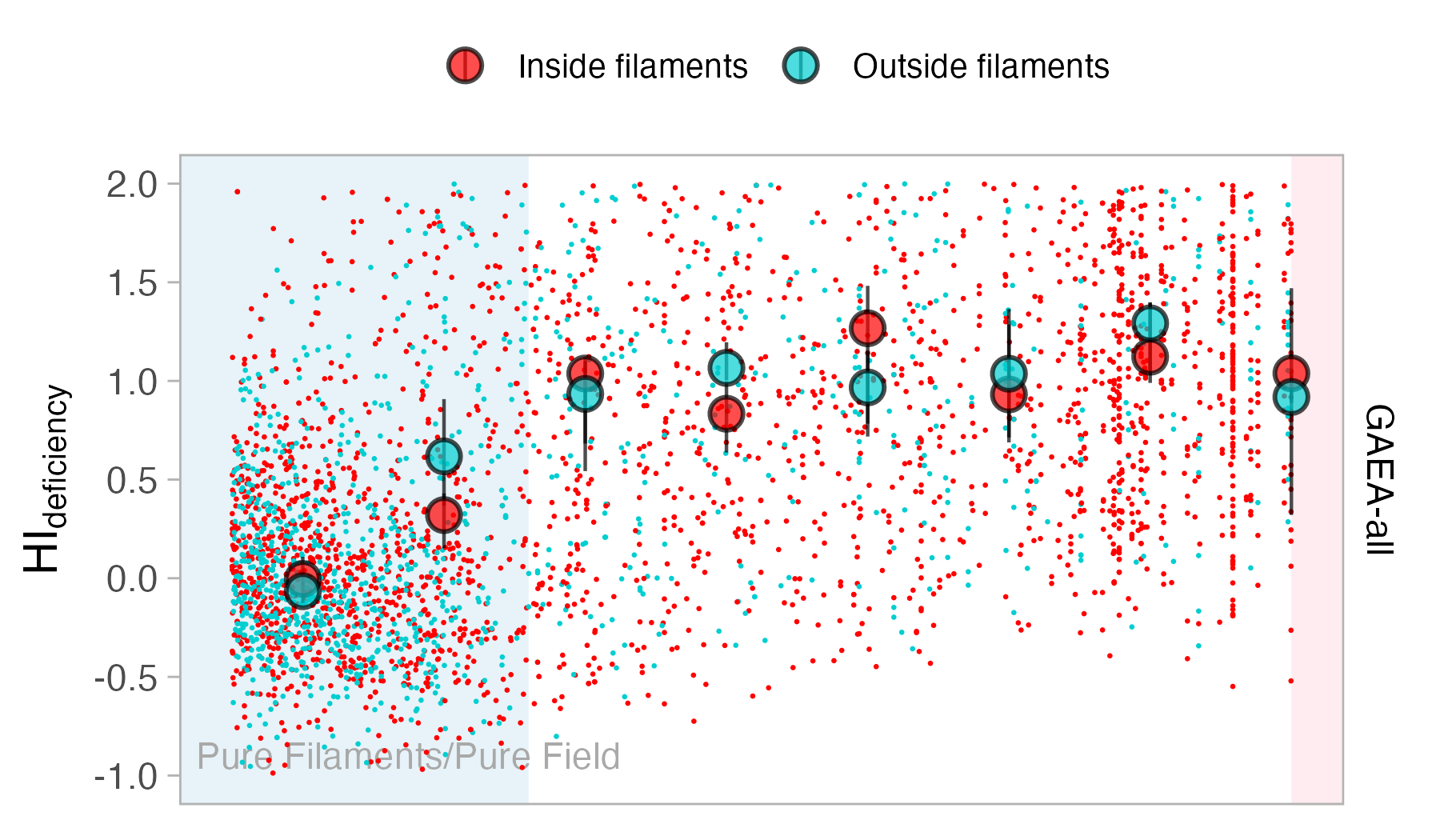}
    \includegraphics[width=1\linewidth]{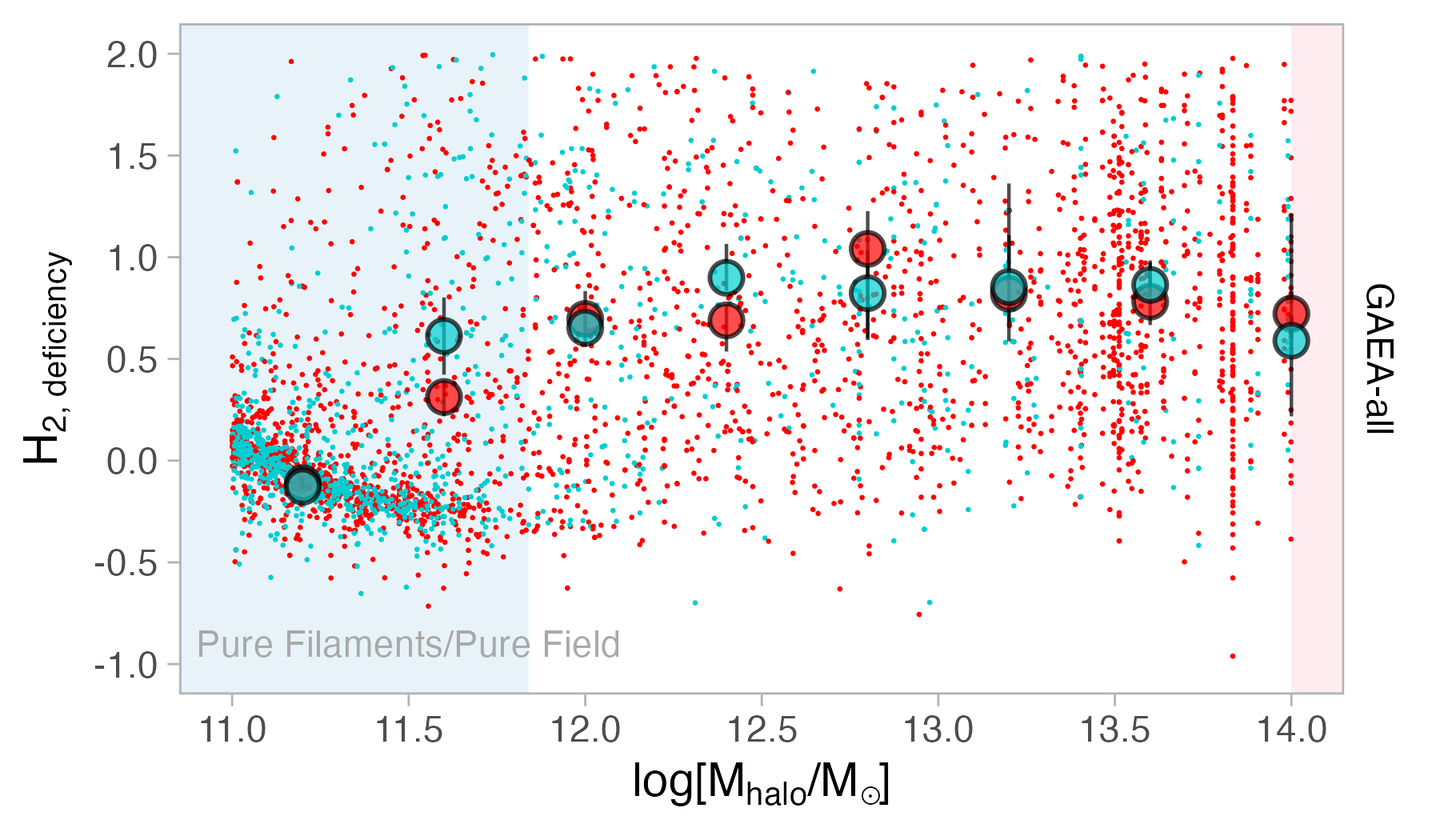}
    \caption{Same as Fig.~\ref{fig:mhidef_mhalo} but with the classification of galaxies inside or outside filaments performed according to the FS identified without consideration of line-of-sight effects.  }
    \label{app:fig:mhalotest_distored}
\end{figure}

\par
\textcolor{black}{Figure~\ref{app:fig:disttest_distored} shows the comparison between \ce{HI}-/\ce{H2}-deficiency profiles for the distorted and `true` filaments. Focusing on filament galaxies, in the case of the distorted FS  both profiles have a rather linear decrease with increasing distance from the filament spine; when considering instead the true filaments, the decrease follows an exponential decline. However, overall this does not affect the main results: a dependency of \ce{HI} or \ce{H2} deficiency as a function of distance from the filament spine is recovered only for filaments and does not exist for pure filament. }

\begin{figure*}
    \centering
    \includegraphics[width=1\linewidth]{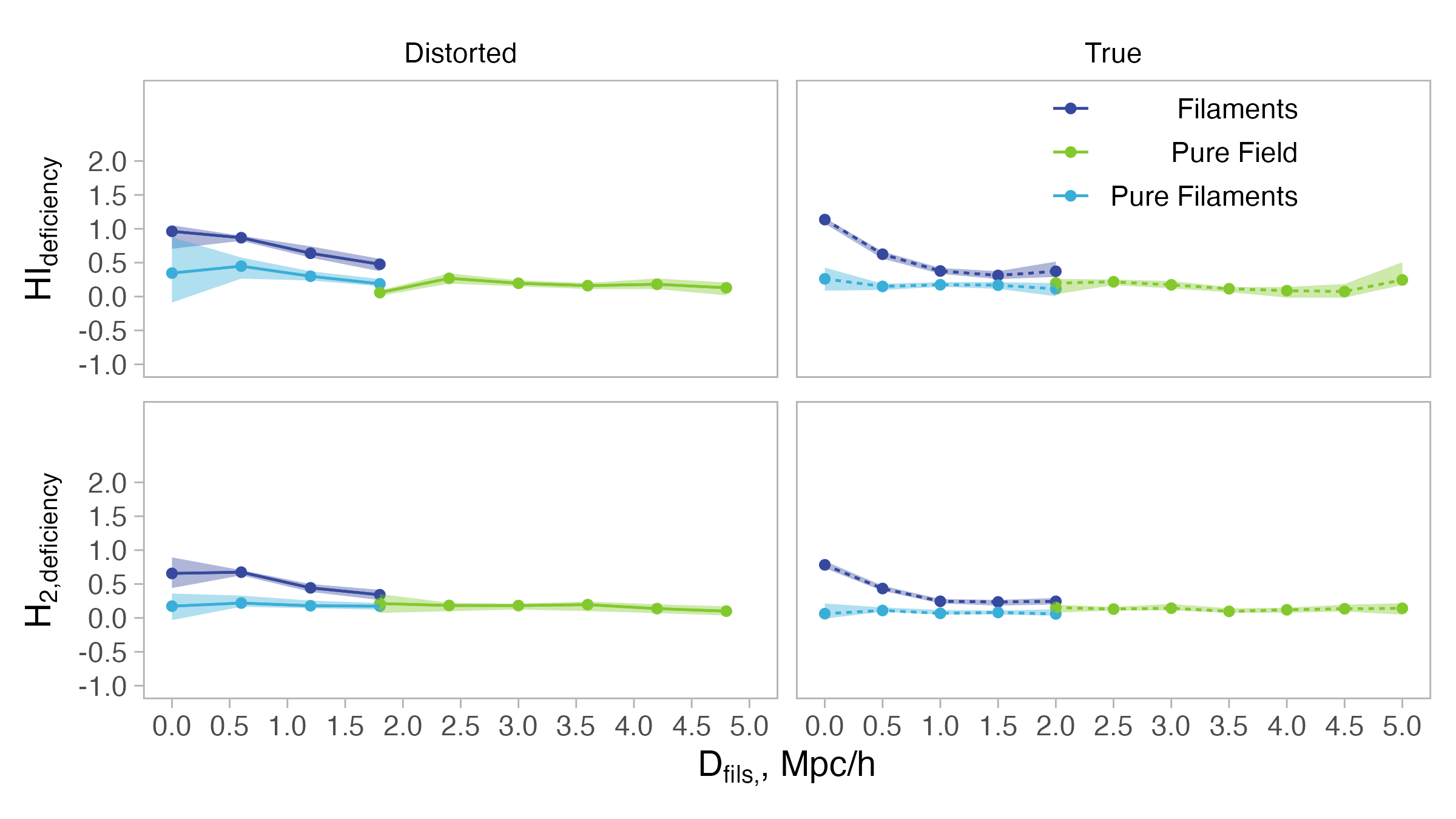}
    \caption{\textcolor{black}{\ce{HI}- or \ce{H2} deficiency (median values and 1$\sigma$ significance interval) as a function of 3D distance to the nearest filament including  line-of-sight distortions~ (left panels, full analogue of Fig.~\ref{fig:mhidef_mh2def_disttoggfs_distored}) and excluding them ~(right panels).}  }
    \label{app:fig:disttest_distored}
\end{figure*}
\end{appendix}
\end{document}